\documentclass[pra,showpacs,reprint,amsmath,amssymb,superscriptaddress,longbibliography,floatfix]{revtex4-1}
\usepackage[utf8]{inputenc}
\usepackage{amsfonts,amssymb,amsmath}

\DeclareUnicodeCharacter{2212}{-}

\usepackage{graphicx}
\usepackage{dcolumn}
\usepackage{bm}
\usepackage{array}

\usepackage{epstopdf}
\usepackage{multirow}
\usepackage{color}
\usepackage{enumerate}
\usepackage{makecell}

\usepackage[caption = false]{subfig}

\usepackage{hyperref}
\usepackage{booktabs}

\begin{document}

\title{Subsystem surface and compass code sensitivities to non-identical infidelity distributions on heavy-hex lattice}

\author{M.~Carroll}
\author{J.~R.~Wootton}
\author{A.~W.~Cross}

\begin{abstract}
Logical qubits encoded into a quantum code exhibit improved error rates when the physical error rates are sufficiently low, below the pseudothreshold. Logical error rates and pseudothresholds can be estimated for specific circuits and noise models, and these estimates provide approximate goals for qubit performance. However, estimates often assume uniform error rates, while real devices have static and/or dynamic distributions of non-identical error rates and may exhibit outliers. These distributions make it more challenging to evaluate, compare, and rank the expected performance of quantum processors. We numerically investigate how the logical error rate depends on parameters of the noise distribution for the subsystem surface code and the compass code on a subdivided hexagonal lattice. Three notable observations are found: (1) the average logical error rate depends on the average of the physical qubit infidelity distribution without sensitivity to higher moments (e.g., variance or outliers) for a wide parameter range; (2) the logical error rate saturates as errors increase at one or a few 'bad' locations; and (3) a decoder that is aware of location specific error rates modestly improves the logical error rate. We discuss the implications of these results in the context of several different practical sources of outliers and non-uniform qubit error rates.
\end{abstract}

\maketitle

\section{Introduction}
\label{Intro}
Quantum computing promises computational speed up in numerous special purpose applications \cite{reiher_elucidating_2017,childs_toward_2018, bravyi_future_2022}. An outstanding challenge is that noisy physical operations limit the depth of reliable quantum circuits. Quantum error correction (QEC) provides a fault-tolerant approach to construct deep, reliable circuits and is believed to be an essential tool for scaling \cite{dennis_topological_2002,fowler_surface_2012,bravyi_future_2022}.
 
 A wide range of qubit gate infidelity distributions are observed in practical devices \cite{iolius_performance_2022, hertzberg_laser-annealing_2021}. A general concern is how a distribution affects logical qubit performance. More specifically, predicting relative performance of sets of qubits with different infidelity distributions is of practical importance for steps like screening selection and acceptance criteria (e.g., quantifying yield and predicting whether outlier infidelities will be tolerable). In this context the qubit set would be an appropriately connected set to implement a quantum error correction code and for which the infidelity distribution was obtained directly or estimated by indirect means \cite{Zhang_LaserAnneal_22}. Recent work has begun to address independent and non-independent, non-identically distributed noise  \cite{hanks_decoding_2020,iolius_performance_2022,clader_impact_2021,berke_transmon_2022}, as well as the existence of thresholds in the presence of inoperable qubits and gates \cite{Strikis_PhysRevApplied23,nagayama_surface_2017,fowler_low_2018,auger_fault-tolerance_2017}.

\begin{figure}[ht!] %h b,t ! 
    \centering
    
    \subfloat[\label{fig:Fig1:a} Sample of measured $T_1$s]{
    	\includegraphics[width=0.23\textwidth, clip,trim= 0mm 0mm 0mm 0mm]{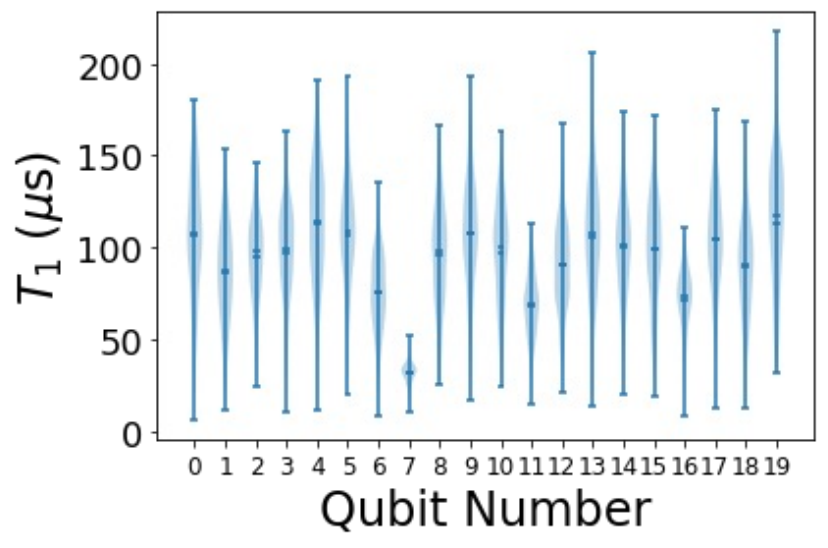}
    }
    \subfloat[\label{fig:Fig1:b} Heavy hex layout]{
        \includegraphics[width=0.29\textwidth, clip,trim= 0mm 0mm 0mm 0mm]{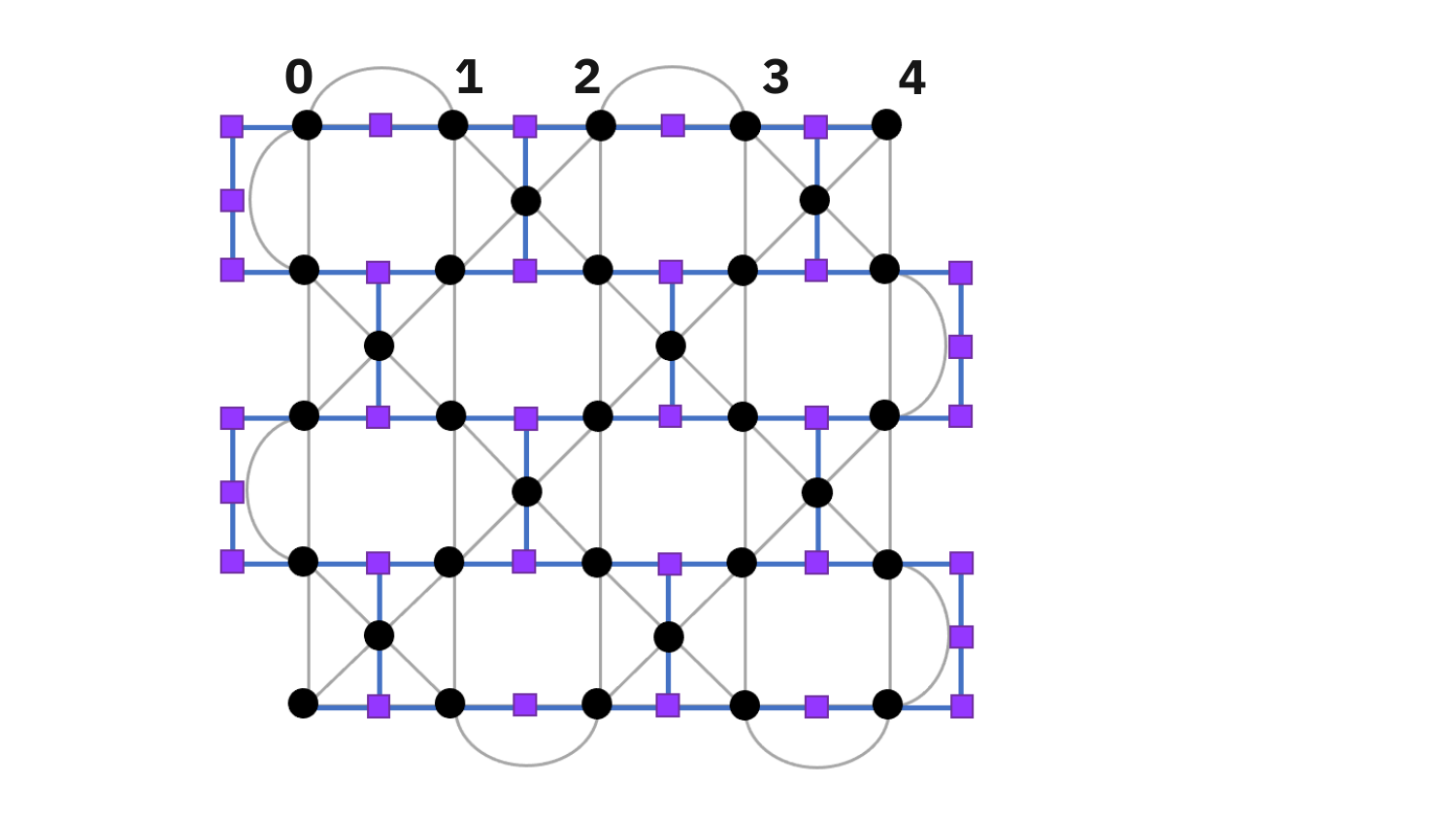}
    }
    \caption{(a) Violin plots of an illustrative distribution of $T_1$ measurements of 20 qubits for which other related statistics of individual qubits in the device were previously published \cite{carroll_dynamics_2022}. (b) Schematic of the rotated subsystem surface code (RSSC) for a distance 5 layout. The qubit types are distinguished by color and shape. Data qubits are black circles and measurement qubits are purple squares.}
    \label{fig:Fig1}
\end{figure}

To provide insight about how logical error rates depend on changing parameters of different distributions of physical qubit error rates, we numerically simulate the \textit{sensitivity} of logical error rates to changes in parameters of several important types of non-identical independent distributed noise. Here we define \textit{sensitivity} as the change in logical error to any change in a parameter of a noise distribution. The numerics show trends that provide qualitative guidance about how to \textit{rank} sets of qubits intended for QEC codes such as surface code. \textit{Ranking} of sets of qubits with non-uniform gate infidelity distributions is in the context of predicting a relative ordering of logical error rates for the different sets of qubits (e.g., predicting which part of a chip will perform a small code better). In the context of providing practical guidance, we choose to examine circuit-level simulations of surface codes and compass codes \cite{licompass} mapped to the heavy hex lattice\cite{LowDegree20}. The heavy hex lattice is chosen, in part, because of its immediate utility in presently available devices \cite{sundaresan_matching_2022}. To provide additional experimental context, we discuss the relationship of these non-identical error distributions to forms of decoherence such as energy relaxation in superconducting qubits, $T_1$, which are known to introduce both spatial and temporal infidelity distributions, Fig. \ref{fig:Fig1} (a) \cite{klimov_fluctuations_2018,carroll_dynamics_2022}.  

We begin with a discussion of the error correction codes used in this work. This includes the first presentation of a rotated subsystem code mapping to the heavy hex lattice and an improved schedule for the heavy hex code \cite{LowDegree20}, section \ref{Codes}. We then introduce the different error rate distribution cases used in this work, section \ref{NoiseModel}. Key results are highlighted for each of the cases in section \ref{resultsSec}, while supporting details are placed in supporting appendices. We provide some perspective on the meaning of these results for areas such as pseudothresholds, screening and modularity in the discussion, section \ref{discussSec} followed by a brief conclusion.

\section{Quantum error correction on a heavy hexagonal layout} \label{Codes}

Control and fabrication constraints can impact the yield of large quantum computing devices based on fixed-frequency transmon qubits \cite{hertzberg_laser-annealing_2021}. This has led to devices with reduced qubit connectivity, using so-called ``heavy'' (or subdivided) lattices where qubits are placed on vertices and edges of a low-degree planar graph (see Fig.~\ref{fig:Fig1}b). The lattice's reduced degree of connectivity eases physical implementation \cite{hertzberg_laser-annealing_2021,Zhang_LaserAnneal_22}.

There is interest to design fault-tolerant operations adapted to these constraints. For example, Chamberland et al. proposed flag error correction circuits for surface codes and compass codes on heavy lattices \cite{LowDegree20}. Here we consider two codes adapted specifically to the heavy hexagonal lattice, a compass code called the heavy hexagon code (HHC) \cite{LowDegree20} and the subsystem surface code (SSC) \cite{Bravyi2013}.

\begin{figure*}[t!]%[htbp]
    \centering
    \subfloat[HHC gauge operators \label{fig:codes:a}]{
	\includegraphics[width=0.2\textwidth]{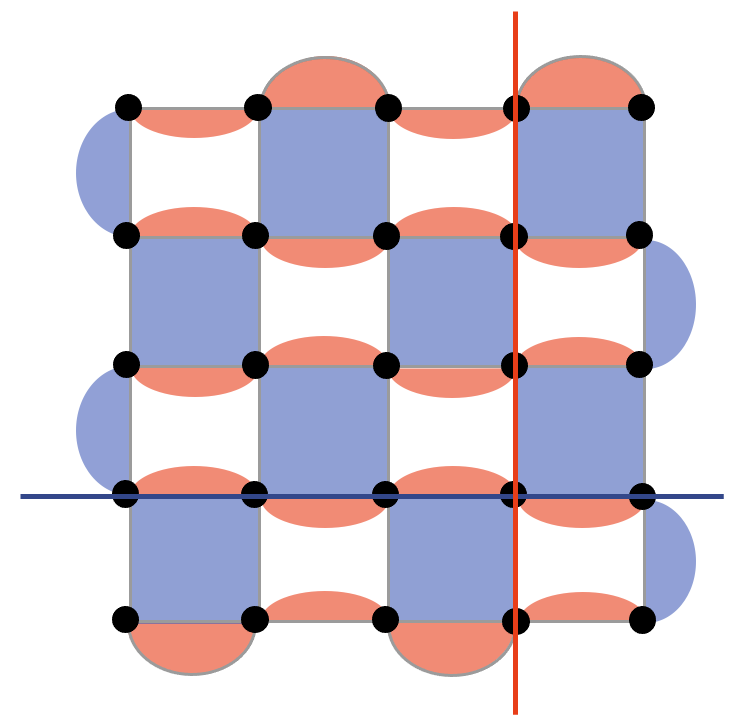}
    } \qquad
    \subfloat[HHC stabilizers \label{fig:codes:b}]{
	\includegraphics[width=0.2\textwidth]{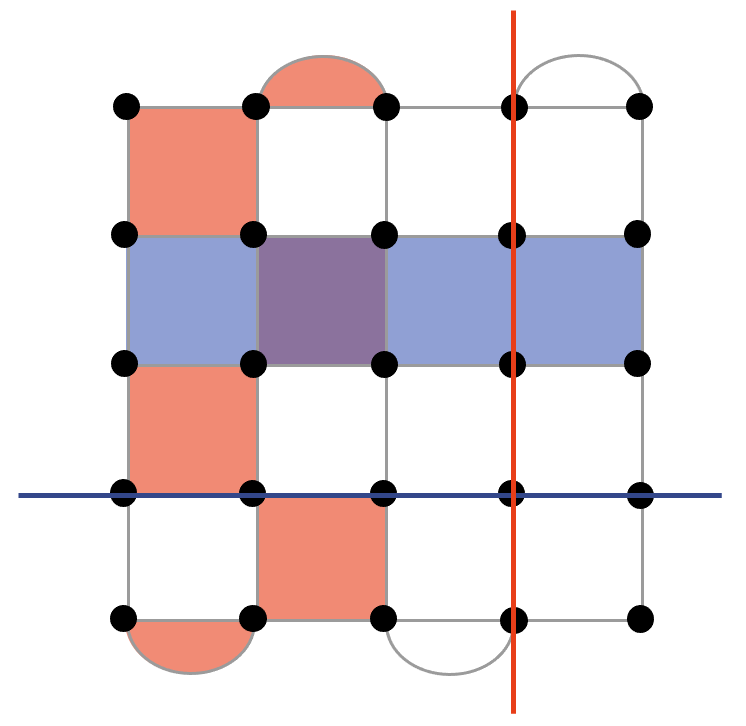}
    } \qquad
    \subfloat[RSSC gauge operators \label{fig:codes:c}]{
	\includegraphics[width=0.2\textwidth]{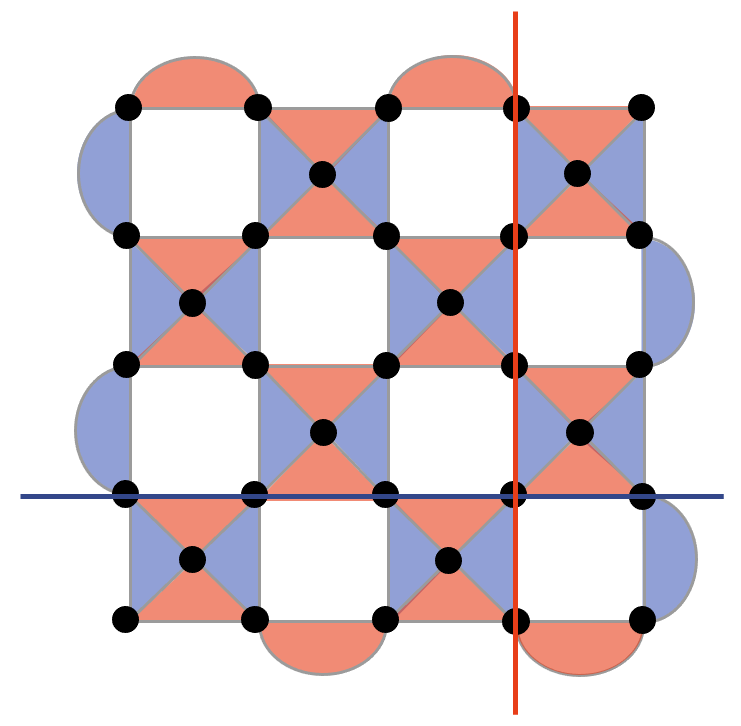}
    } \qquad
    \subfloat[RSSC stabilizers \label{fig:codes:d}]{
	\includegraphics[width=0.2\textwidth]{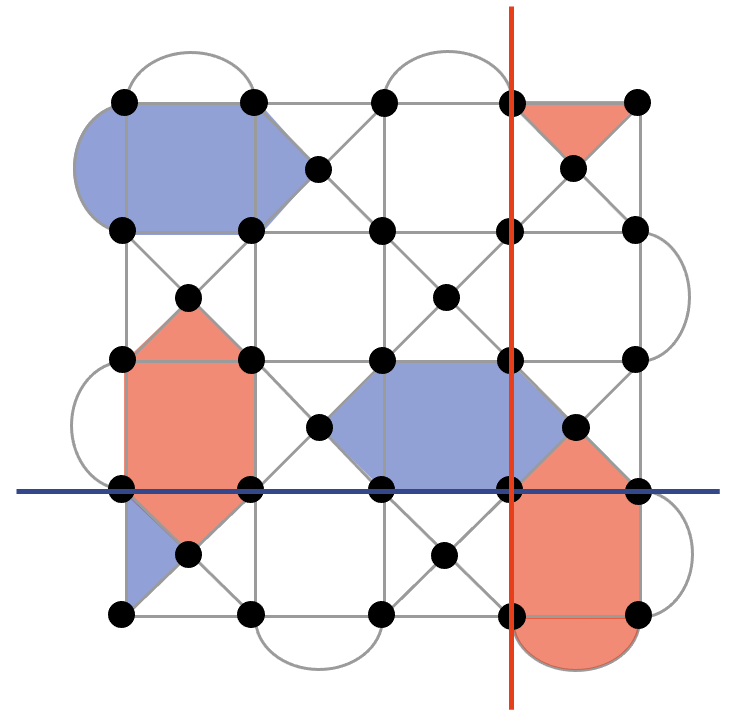}
    }
    \caption{\textbf{Heavy-hexagon code (HHC) and rotated subsystem surface code (RSSC).} Data qubits are placed on vertices of each graph. Pauli Z (X) operators are shown in blue (red), and logical operators are indicated by lines. (a,b) HHC has weight-2 and 4 gauge operators, weight-$2d$ Bacon-Shor stabilizers, and weight-4 surface code bulk stabilizers. (c,d) RSSC has weight-3 (triangular) bulk gauge operators and weight-6 (hexagonal) bulk stabilizers.}
	\label{fig:codes}
\end{figure*}

\subsection{Heavy hexagon code (HHC)}

The heavy hexagon code is a subsystem stabilizer code \cite{poulin05} defined by the gauge group
\begin{align}
G = \langle & X_{i,j} X_{i,j+1}, Z_{i,j} Z_{i,j+1} Z_{i+1,j} Z_{i+1,j+1}, \\ \nonumber
& Z_{2m−1,1} Z_{2m,1}, Z_{2m,d} Z_{2m+1,d}\rangle
\end{align}
where $i,j=1,2,\dots,d-1$, $m=1,2,\dots,\frac{d-1}{2}$, and $i+j$ is odd for the second term, as shown in Fig.~\ref{fig:codes:a}. We choose odd distances $d$ throughout this paper. The corresponding stabilizer group, illustrated in Fig.~\ref{fig:codes:b}, is
\begin{align}
S = \langle & X_{i,j} X_{i,j+1} X_{i+1,j} X_{i+1,j+1}, X_{1,2m} X_{1,2m+1}, \\ \nonumber
& X_{d,2m-1} X_{d,2m}, \prod_{j} Z_{i,j} Z_{i+1,j}\rangle
\end{align}
with $i+j$ even for the first term. The X stabilizers are the same as the surface code whereas the Z stabilizers are those of the Bacon-Shor code. This code has a threshold of $p_{th}\sim 0.0045$ on the heavy hexagonal lattice for errors detected by the surface code stabilizers, and although there is no corresponding threshold for the Bacon-Shor code stabilizers, low logical error rates can still be achieved \cite{LowDegree20}. Quantum error correction has been experimentally demonstrated on the $d=3$ HHC \cite{Sundaresan_2023}.

In this work, we apply the idea of schedule-induced gauge-fixing \cite{higgott_subsystem_PhysRevX} to optimize the total logical error probability of the HHC. This is a general idea that can be applied to CSS subsystem codes wherein gauge operators with deterministic eigenvalues are treated as stabilizers during syndrome processing.
For the HHC, the circuits for each gauge operator measurement round are identical to \cite{Sundaresan_2023}, but the schedule may now repeat the same set of gauge operator measurements multiple times (see Appendix~\ref{appendix:RSSC} for details). The decoding algorithm is also identical to the matching decoder in \cite{Sundaresan_2023}, but detection events are defined with respect to the instantaneous stabilizer groups as described in \cite{higgott_subsystem_PhysRevX}.

\subsection{Rotated subsystem surface code (RSSC)}

The subsystem surface code (SSC) \cite{Bravyi2013} and the standard surface code are related to each other by a local, constant depth stabilizer circuit. However, the gauge operators of the SSC have weight-3 or less and therefore can be measured by circuits that are naturally fault-tolerant. Higgott and Breuckmann have shown that subsystem surface codes can have high thresholds of $0.85\%$ under circuit model depolarizing noise \cite{higgott_subsystem_PhysRevX}, which exceeds the $0.67\%$ threshold of the surface code.

We focus on a rotated form of the SSC \cite{brown19} that is defined by the gauge group
\begin{align}
    G = \langle & Z_{i,j} Z_{i,j+1} Z_{t(i,j)}, Z_{i+1,j} Z_{i+1,j+1} Z_{t(i,j)}, \\
    & Z_{2m−1,1} Z_{2m,1}, Z_{2m,d} Z_{2m+1,d}, \\
    & X_{i,j} X_{i+1,j} X_{t(i,j)}, X_{i,j+1} X_{i+1,j+1} X_{t(i,j)}, \\
    & X_{1,2m-1} X_{1,2m}, X_{d,2m} X_{d,2m+1} \rangle,
\end{align}
where $i,j=1,2,\dots,d-1$, $m=1,2,\dots,\frac{d-1}{2}$, and $i+j$ is odd for the weight-3 operators, as shown in Fig.~\ref{fig:codes:c}. The index $t(i,j)$ refers to the qubit in the center of the face with corners $(i,j)$, $(i+1,j)$, $(i,j+1)$, $(i+1,j+1)$, where $i+j$ is odd. A distance $d$ code has $(d-1)$ weight-2 X (Z) gauge operators and $(d-1)^2$ weight-3 X (Z) gauge operators. We use odd distances $d$ throughout the paper. The corresponding stabilizer group is
\begin{align}
    S = \langle & Z_{t(i,j-1)} Z_{i,j} Z_{i+1,j} Z_{i,j+1} Z_{i+1,j+1} Z_{t(i,j+1)}, \\
    & X_{t(i-1,j)} X_{i,j} X_{i,j+1} X_{i+1,j} X_{i+1,j+1} X_{t(i+1,j)} \rangle,
\end{align}
where $i,j=1,2,\dots,d$ with $i+j$ is even. The operators $X_{i,j}$, $Z_{i,j}$, $X_{t(i,j)}$, and $Z_{t(i,j)}$ are defined to be identity whenever $i,j\notin \{1,2,\dots,d\}$. Example stabilizers, which are hexagonal in the bulk, are drawn in Fig.~\ref{fig:codes:d}. Since each pair of X and Z stabilizers that overlaps on four qubits has an associated pair of independent, anticommuting gauge operators, an RSSC with odd minimum distance $d$ encodes one logical qubit and $(d-1)^2/2$ gauge qubits into $d^2 + (d-1)^2/2$ data qubits.

The RSSC is a promising option for implementing fault-tolerant quantum computing on a heavy hexagonal lattice for several reasons. First, the gauge operators can be measured by simple circuits. Second, the RSSC has a threshold, so it is expected to asymptotically out-perform the compass code. Finally, techniques for fault-tolerant computing with the surface code carry over to the subsystem surface code and differ only in implementation details.

The RSSC naturally overlays the heavy hexagonal lattice as shown in Fig.~\ref{fig:Fig1:b}. We refer to extra qubits for making or mediating measurements as measurement qubits. Each X gauge operator uses a single measurement qubit on the interior, and the weight-2 Z gauge operators on the boundary use three measurement qubits, assuming translation symmetry. Counting these additional $d^2+2d-3$ ancillary qubits, we operate a distance-$d$ code using a subset of the lattice containing $5d^2/2+d-5/2$ qubits.

\begin{figure}[h]
\centering
\subfloat[X-type gauge operators \label{fig:gauge-circuit:a}]{
    %\begin{tikzpicture}  % keep this source code
    %\begin{yquant}
    %qubit q[3];
    %qubit {$|+\rangle$} a[1];
    %cnot q[0] | a;
    %cnot q[1] | a;
    %cnot q[2] | a;
    %h a;
    %measure a;
    %\end{yquant}
    %\end{tikzpicture}
    \includegraphics[width=0.2\textwidth]{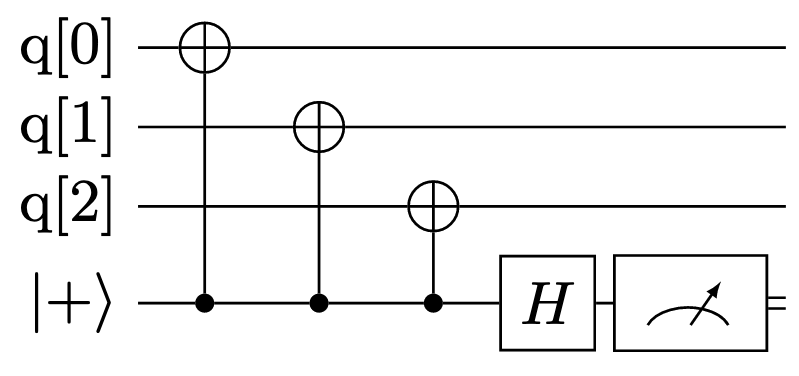}
} \qquad
\subfloat[Z-type gauge operators \label{fig:gauge-circuit:b}]{
    %\begin{tikzpicture}  % keep this source code
    %\begin{yquant}
    %qubit {$|q_0\rangle$} q;
    %qubit {$|0\rangle$} a;
    %qubit {$|q_1\rangle$} q[+1];
    %qubit {$|0\rangle$} a[+1];
    %qubit {$|q_2\rangle$} q[+1];
    %cnot a[0] | q[0];
    %cnot a[1] | q[2];
    %cnot q[1] | a[1];
    %cnot a[0] | q[1];
    %cnot q[1] | a[1];
    %cnot a[1] | q[2];
    %measure a[0];
    %\end{yquant}
    %\end{tikzpicture}
    \includegraphics[width=0.2\textwidth]{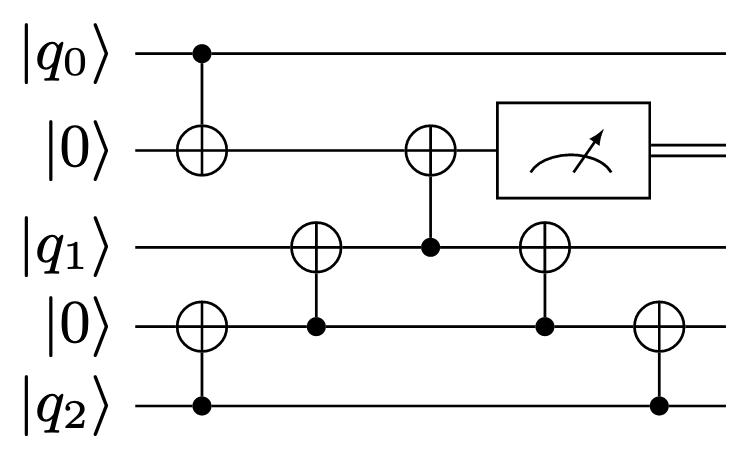}
}
\caption{Circuits for measuring gauge operators of the subsystem surface code on the heavy-hexagon lattice}
\label{fig:gauge-circuit}
\end{figure}

Due to the natural mapping to the heavy hexagon lattice, the X gauge operators can be measured without extra gates using the circuit in Fig. \ref{fig:gauge-circuit:a}. Bit-flip errors on the measurement qubit propagate to at most one bit-flip error on the data modulo the gauge group, so there are no space-like hook errors. However the circuits can produce space-time hook errors from a single fault event, just like the surface code.

The Z gauge operators are more difficult to measure but each can be measured using the circuit in Fig.~\ref{fig:gauge-circuit:b}. By inspecting all single-fault events, we conclude that this circuit has the property that any single fault gives rise to at most one X and one Z error on the data modulo the gauge group. Since these are correctable errors for CSS codes, the syndrome measurement circuit is fault-tolerant in the following sense. If $w$ faults occur then the output state deviates by at most $w$ X errors and $w$ Z errors modulo the gauge group. These X and Z errors can be corrected independently as long as there is no mechanism to convert them into Y errors.

On the heavy hexagon lattice, we find that the RSSC has a threshold of nearly $0.3\%$ for circuit-level noise. Additional details and a comparison with the HHC can be found in Appendix~\ref{appendix:RSSC}.

\subsection{Other relevant codes}

It is worth mentioning other codes that can be directly implemented on the heavy hex layout, which are not included in this study. In particular, Floquet codes~\cite{hastings_haah_2021}, Floquet color codes~\cite{kesselring_anyon_2022}, matching codes~\cite{wootton_matching_2022} and repetition codes~\cite{liepelt_arc_2023}. In all these cases, syndrome measurements are made via two-qubit parity operators, for which the qubits on the edges of the heavy hex lattice can be used as auxiliaries.

Repetition codes are very much an outlier in this group since they protect only a logical bit rather than a logical qubit, and have a distance equal to the number of data qubits. We therefore do not include them in our study, since the results are likely be unrepresentative of truly quantum codes.

For Floquet codes, Floquet color codes and matching codes, though a direct implementation of these is possible on the heavy hex lattice, they are more ideally suited to systems in which direct parity measurements are possible~\cite{paetznick_majorana_2023}. Compared to this, implementation with the heavy hex would result in a significant decrease in threshold and performance, and significantly less resilience when there is a bias towards different noise sources~\cite{hetenyi_tailoring_2023}. It is for this reason that we do not include such codes in this study.

\section{Noise models and methods}
\label{NoiseModel}
In this section we introduce general noise models used in this work. For all models, the circuits are decomposed into a set of operation types, $\mathcal{C}$ = \{cx,h,s,id,x,y,z,measure,initialize,reset\}, the first seven abbreviated operation types being the cnot, hadamard, phase, identity, $x, y, z$ rotations, respectively. For iid, each of the operation types experiences a fault with a depolarizing noise error rate specific to the operation type. Non-identical, independent, distribution (niid) assignments used in this work will be further detailed below. Stabilizer simulations of the noisy circuits are done using the Qiskit software stack. 

Decoding is done using minimum weight perfect matching (MWPM) \cite{dennis_topological_2002}. The decoder is provided with either all location specific error rate information, called \textit{aware} condition, or limited to using the lattice averages for each operation type, called \textit{naive}. Briefly the MWPM method forms a decoding hypergraph of nodes representing Z or X stabilizer measurements. Edges of the graph are weighted according to the available error rate information. A perfect matching algorithm returns a minimum weight set of connecting edges for each syndrome of stabilizer measurements at the nodes that highlight the presence of an error.  \textit{Aware} decoding is used unless otherwise noted. Further details of the simulation method can be found in Sundaresan et al. \cite{Sundaresan_2023}. 

We simulate logical error rates, $\langle p_{out} \rangle$, for cases composed of different conditions defined from: (i) type of lattice distribution of error rates (i.e., noise model), (ii) code, (iii) decoder type (\textit{aware} or \textit{unaware}), and (iv) which distribution parameter is varied, Table \ref{table:modelSec}. 

We label the distributions categories as: \textit{uniform}, \textit{normal}, \textit{non-normal} and \textit{location specific}. The \textit{uniform} signals that error rates for the faulty circuit operation, $cx, h$ and $id$ (unless otherwise noted), are set to the same input error rate, $p_{in}$, at all lattice locations. The errors are modeled as depolarizing noise. A background error rate, typically negligibly small compared to the $p_{in}$ is used for the other faulty circuit operations for these test cases unless otherwise noted. 

The \textit{normal} distribution signals that each faulty $cx, h$ and $id$ location is assigned an error rate drawn from the absolute value of a normal distribution (unless otherwise noted). The normal distribution is defined with an average error rate, $\langle p_{in} \rangle$ and a standard deviation $\sigma = \alpha \langle p_{in} \rangle$. The errors are again modeled as depolarizing noise. Two steps of averages are done to obtain an output logical error rate. In the first step, the results are averaged for a single device instance of unique fixed error rates at each faulty location. Each \textit{shot} of a circuit includes $d$ rounds of zzxx(xxzz) stabilizer measurements for X(Z) intialization and measurement, respectively. For the second step, an average is formed over a number of device \textit{instance}s that we define as a new distribution of error rates from the error distribution category.

A \textit{non-normal}, reciprocal-normal distribution is considered in this work that is a proxy, for example, for coherence limited errors and more generally distributions with heavy tails at higher error rates, see appendix \ref{appendix:infdist_T1} and \ref{appendix:RecipOfNormDist}. We also examine cases where one to four locations have error rates that are either fixed at $p_{in}=0.5$ or varied between $\langle p_{in} \rangle < p_{in} < 0.5$, labeled \textit{location} (e.g., 'bad' sites). An errant location assigns increased error rates for the operations $h, id$ and the two qubit operations, $cx$, connected to the qubit site. The rest of the lattice error rates are constant, $\langle p_{in} \rangle$. Further details for each case are indicated in the results sections.

\begin{table}[ht!]
\begin{tabular}{||c | c | c | c |c |c ||} 
\hline
\makecell{ Variable } & \makecell{$p_{in}$} & \makecell{$\sigma$} & \makecell{N loc.} & \makecell{Code}&\makecell{Decode} \\
 \textbackslash Distribution  & & & & & comp.  \\
\hline\hline
Uniform & S.\ref{Codes},A.\ref{appendix:RSSC} & -  & - & R,H & - \\
\hline
Normal & - & S.\ref{decodeComp}, A.\ref{appendix:normDistSims} & - & R,H & S.\ref{decodeComp}   \\
\hline
Non-normal & S.\ref{infdist_sigma} & S.\ref{infdist_sigma} & - & R & - \\ 
\hline
Location  & S.\ref{infdist_bad}, A.\ref{Appendix:FuncSites}  & - & S.\ref{infdist_bad}  & R,H & A.\ref{Appendix:FuncSites}  \\ 
\hline
\end{tabular}
\caption{Table of section numbers and codes examined cross referenced to the category of error distribution and parameters varied. The letters S or A indicate section or appendix, respectively, and R or H indicate, RSSC, respectively. - indicates not applicable or unavailable.}
\label{table:modelSec}
\end{table}

\section{Results}
\label{resultsSec}
\subsection{Impact of normally distributed coherence times on logical error rates}
\label{infdist_sigma}

In this section we describe results from a parameterized error distribution for which the error rate is reciprocally related to a normally distributed random variable, $\mathbf{p} \propto 1/ |\mathbf{X}|$. A physical motivation for this distribution is randomly distributed coherence times, like energy relaxation with a coherence time $T_1$. We assign an error rate $\mathbf{p} = (1 - e^{\tau/|\mathbf{T_1}|}) \sim \tau / |\mathbf{T_1}|$, where $\mathbf{T_1}$ is a random variable pulled from a normal distribution with mean, $\langle T_1 \rangle$, standard deviation, $\sigma$ and $\tau$ is a constant representative of an effective gate time of the qubit operation. We simulate output error dependence on the standard deviation of the normal distribution, $\sigma$, for the RSSC code.

The logical output error rate dependence on uniform $p_{in}$ is shown in Fig. \ref{fig:Fig2} (a) (see appendix \ref{appendix:RSSC} for more details) and the dependence on varying $\sigma$ is shown in Fig. \ref{fig:Fig2} (b). We show the dependence normalized to the average and therefore vary a parameter $\alpha = \sigma / \langle T_1 \rangle$. The average error rate in Fig. \ref{fig:Fig2} (b) is $\mu = \tau/\langle T_1 \rangle = 10^{-3}$, an average \textit{cx} error rate foreseeable in the near future ~\cite{Stehlik_PhysRevLett.127.080505}. We show the logical error for both X or Z initialization and measurement cases. The output error rate shows an onset of increasing average logical error rate when $\sigma$ approaches $\alpha \sim 0.2-0.5$.   

We now discuss the numerics in the context of a simple model that illustrates that the onset of increasing $\langle p_{out} \rangle$ may be understood as the onset of the mean of the input error rate, $\langle p_{in} \rangle$ increasing (i.e., $\mu \neq \langle p_{in} \rangle$). We first observe that numerical simulations of logical error rates for normally distributed error rates show no dependence on the standard deviation of the distribution, see appendix \ref{appendix:normDistSims}, in contrast with the $\sigma$ dependence of the reciprocal normal distribution. To understand this observation, we note that a phenomenological model of the repetition code with normally distributed error rates results in an $\langle p_{out}^{rep} \rangle = 3\bar{\epsilon}^2-2\bar{\epsilon}^3$ irrespective of the standard deviation of the error rates, see appendix \ref{appendix:repModel}. Here we define an average output error of the phenomenological model, $\langle p_{out}^{rep} \rangle$, with contributing data sites at indices $i$. Each data site has normally distributed error rates, $\epsilon_i$ and an average error rate of all the data, $\bar{\epsilon}$. The phenomenological model provides the insight that random distributed error rates of a faulty type of component will result in an averaged contribution to the output error rate consistent with the numeric result. We conjecture that for a wide range of uncorrelated error rates the average logical error is a function of the average physical qubit error rate, $\langle p_{out} \rangle = F(\langle p_{in} \rangle)$, in contrast to dependence on higher moments of the distribution. On the other hand, we also note that although $\langle p_{out}^{rep} \rangle$ is not \textit{sensitive} to changes in $\sigma_{\epsilon}$, the 'device-to-device' variance of $\langle p_{out}^{rep} \rangle$ is predicted to be dependent on $\sigma_{\epsilon}$, see appendix \ref{appendix:repModel}. 

We return to the observation that the logical error rate increases for $\alpha \sim 0.2-0.3$ when the error rate distribution of the faulty locations is proportional to $\tau/|\mathbf{T_1}|$. The mean of such a distribution depends on the harmonic mean of the $\mathbf{T_1}$ distribution. The average is done for a normally distributed $T_1$ distribution, $P(\mathbf{T_1}=T_1) = \frac{1}{\sqrt{2\pi\sigma}}e^{\frac{-(T_1 - \langle T_1 \rangle)^2}{2\sigma^2}}$, see appendix \ref{appendix:RecipOfNormDist} for more details. The $T_1$ values sampled above versus below $\langle T_1 \rangle$ are unevenly weighted and this leads to a $\sigma$ dependence of $\langle p_{in} \rangle$ despite $\langle T_1 \rangle$ being independent of $\sigma$ consequently leading to an increase in $\langle p_{out} \rangle$.

This behavior is of general interest for any error process that depends reciprocally on a random process such as electronics noise impacting dephasing (e.g., $T_{2}$), although relative weights in the tails could lead to a deviation from the simple picture that the output error rate can be predicted by the average input error rates as if they were uniform error rates. Effects of outliers will be discussed in the following section.  

\begin{figure*}[t] %h b,t ! 
    \centering
    \subfloat[\label{fig:Fig5:a} RSSC $p_{out}$ for uniform $p_{in}$]{
    	\includegraphics[width=0.45\textwidth, clip,trim= 22mm 0mm 22mm 0mm]{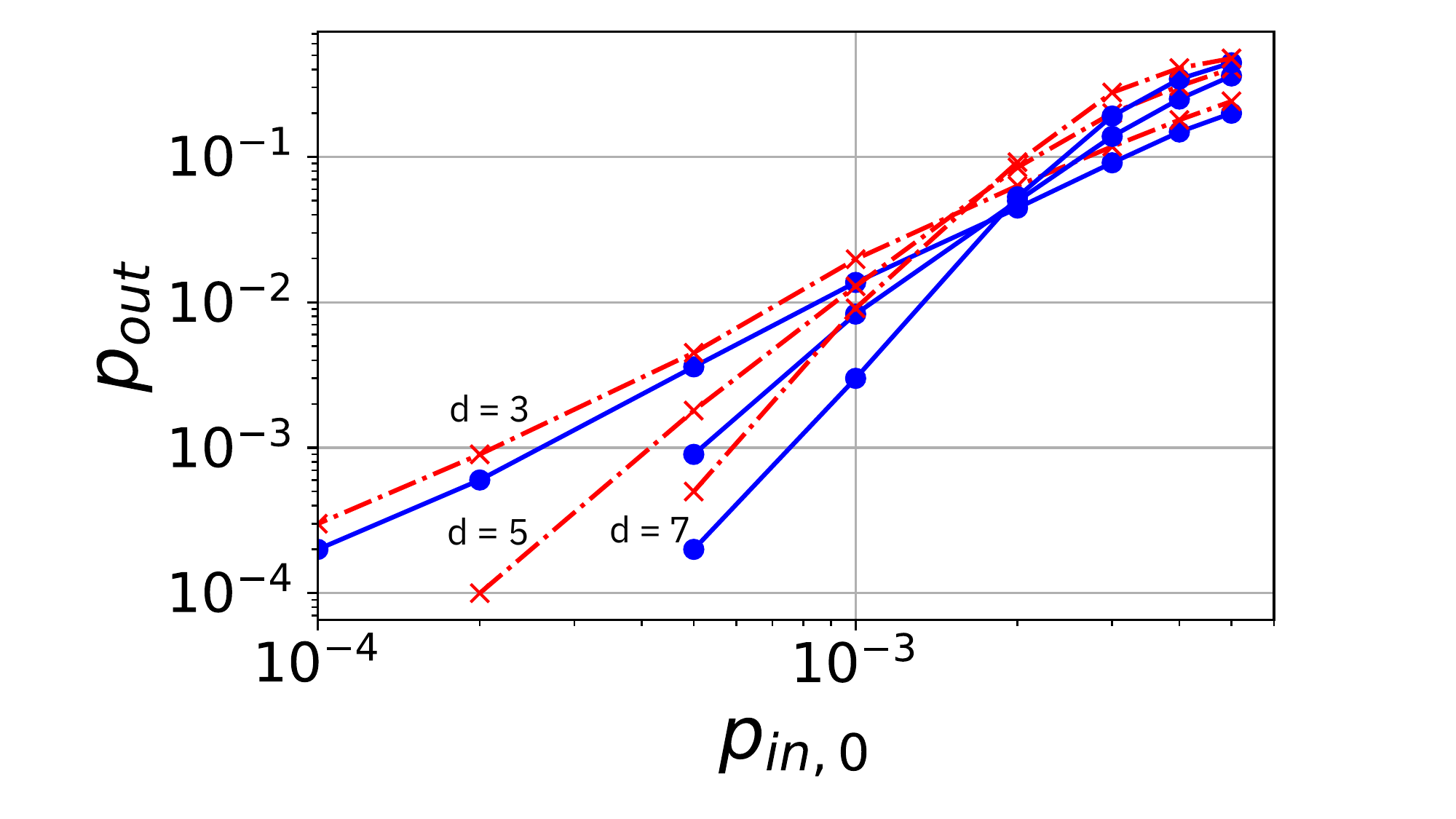}
    }
    \subfloat[\label{fig:Fig5:b} RSSC $p_{out}$ response to varying $\sigma$]{
        \includegraphics[width=0.4\textwidth, clip,trim= 0mm 4.5mm 5mm 5.5mm]{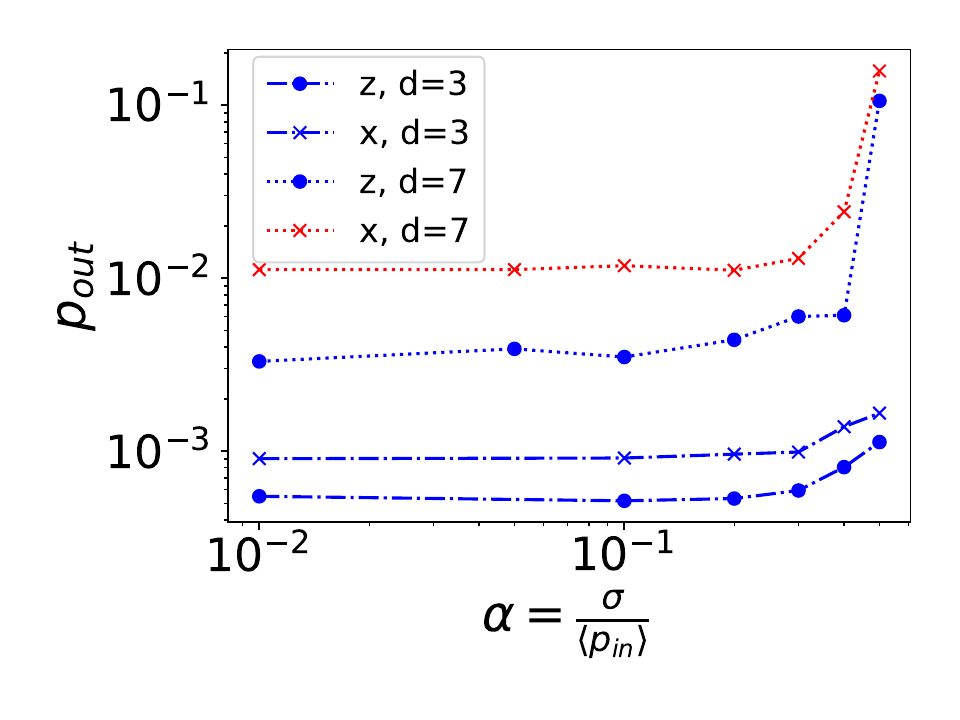}
    }
    
    \caption{(a) Logical output error rates for the RSSC for numerical simulations of distance 3, 5, and 7. The (red) X-measurement and (blue) Z-measurement cases are both shown. (b) Logical output error dependence on an infidelity distribution generated as {\textbf{$ p_{in} = \frac{\tau}{ T_1 }$}} for which {\textbf{$T_{1}$}} is a normally distributed variable. The error rate is sampled uniquely for each operation and held constant throughout the simulation for distance 3 and 7. X initialization, measurement (red) an Z initialization, measurement (blue) bases. The simulations were run for 4, 64 lattice instances, 250k, 10k shots each and $\langle p_{in} \rangle = 2\times10^{-4}, 1\times10^{-3}$, respectively. A single instance for the X and Z basis are compared to the average cases $\langle X \rangle$ and $\langle Z \rangle$. The standard deviation of the distribution is normalized to the mean input error rate (i.e., $\langle p_{in} \rangle = \frac{\tau}{\langle T_{1} \rangle}$, $\sigma = \alpha \langle \epsilon \rangle$).}
    \label{fig:Fig2}
\end{figure*}

\subsection{Dependence of output error on high infidelity outliers at specific qubit sites}
\label{infdist_bad}
In this section we examine the effect of changing error rate at a limited number of sites while the rest of the error rates for faulty locations are held constant. We ask, for example, how \textit{sensitive} the logical error rate is to deviations of error rate at a single faulty location from the average?

We first examine the effect of increasing the error rates, $p_{in,0}$, of \textit{h, id} and all \text{cx} operations that include the data qubit located at site 0, Fig. \ref{fig:Fig1} (a). All other error rates are set to $5\times10^{-4}$. A S-curve behavior is observed as $p_{in,0}$ is increased. While $p_{in,0}$ is similar to $\langle p_{in} \rangle$, little effect is observed on $p_{out}$, consistent with $\langle p_{in} \rangle$ not being significantly changed, see discussion above. At higher $p_{in,0}$, $p_{out}$ begins to rise and then saturates. Rapid increase in output error rate begins at an input error rate of $\sim10^{-2}$ approximately 40 times larger than $\langle p_{in} \rangle$ for $d=5$. A single data qubit location can be relatively faulty compared to the devices average (i.e., $\langle p_{in} \rangle \sim 5\times10^{-4}$) before the output error rate begins to be substantially degraded. The maximum increase in error is bounded further by the commensurate loss in distance of the code.

\begin{figure*}[t] %h b,t ! 
    \centering
    \includegraphics[height=6.5 cm, width=\linewidth, clip,trim= -20mm 22mm -20mm 15mm]{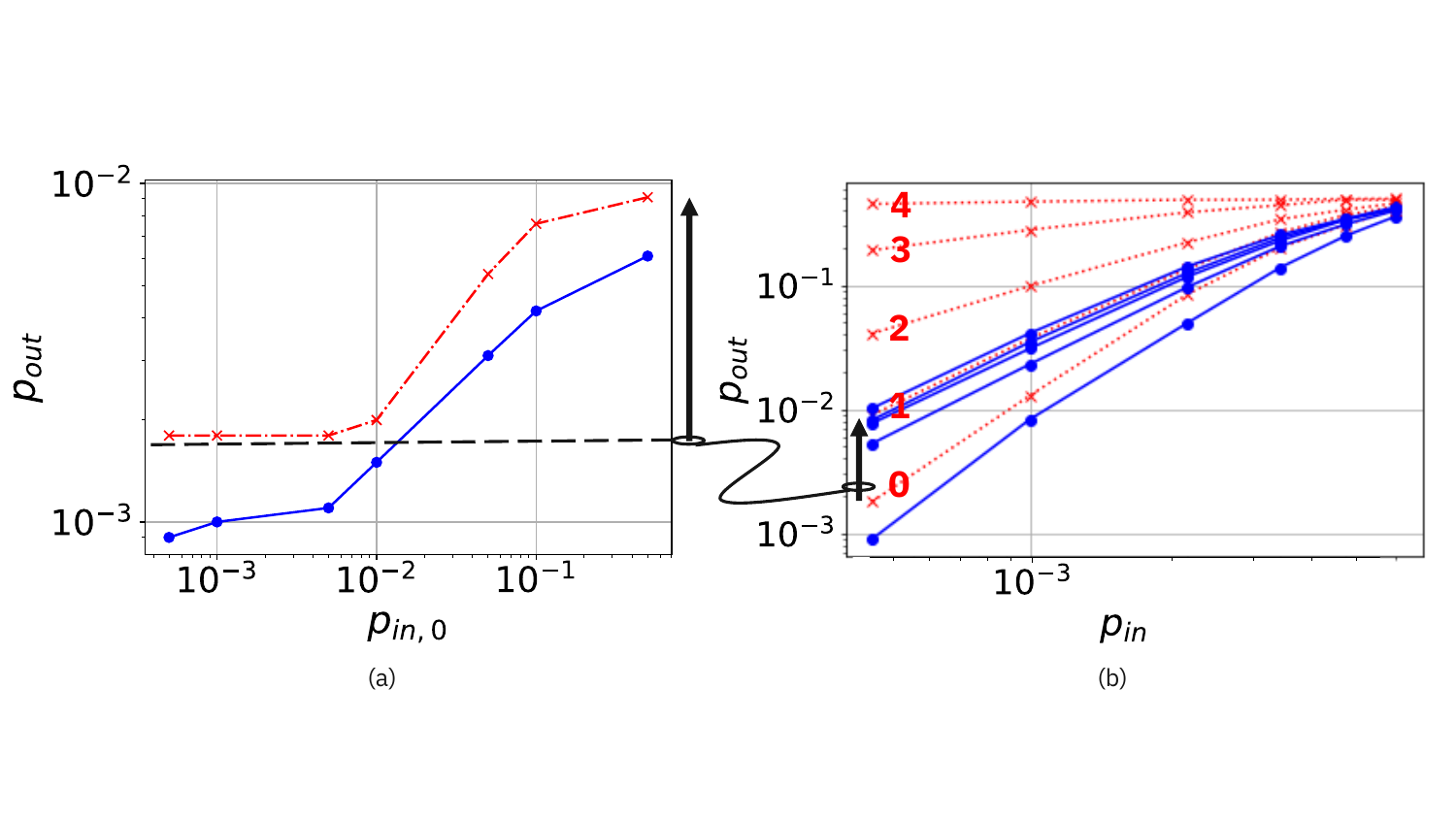}
    \caption{(a) Output error rate dependence on a 'bad' qubit at site 0 (see Fig. \ref{fig:Fig1}(a) for distance 5 using $\langle p_{in} \rangle = 5 \times 10^{-4}$. (b) Output error for distance 5 with $p_{in}=0.5$, 'bad', data qubits at positions 0, 1, 2, 3 and 4 (Fig. \ref{fig:Fig1} (a)) and variable $\langle p_{in} \rangle$ for gate operations associated with qubits sites that do not interact with the 'bad' sites. Two sets of 4 lines are shown. From lowest to increasing output error, the 0 to 4 sites are populated with 'bad' qubits. Red indicates X-measurement and blue indicates Z-measurement.}
    \label{fig:Fig3}
\end{figure*}

To further illustrate the effect of 'knocking out' data qubits, we simulate output error as a function of uniform input error for instances where 0 to 4 data qubits are set to $p_{in}=0.5$, Fig. \ref{fig:Fig3}. We note that the sites 0 to 4 are along the logical Z operator in the RSSC. The measure X output error jumps after the addition of a single 'bad' qubit at site 0 and ouput error increases quasi-linearly with the device $\langle p_{in} \rangle$, Fig. \ref{fig:Fig3} (b). The end points of  the \textit{sensitivity} analysis of output error to varying the $p_{in,i}$ of a single 'bad' qubit at site $i$, Fig. \ref{fig:Fig3} (a), are identifiable in Fig. \ref{fig:Fig3} (b) as indicated. We infer that there are similar S-curves between the other points in Fig. \ref{fig:Fig3} (b). 

Increasing the number of 'bad' data qubits along the Z logical operator leads to monotonic jumps in output error rate. Qualitatively this could be interpreted as the sequential reduction of the effective distance of the code through 'knocking out' code qubits along a logical operator. The 0 site is along both the logical X and Z operators. 'Knocking out' the 0 site with a 'bad' qubit reduces the correcting power of the logical qubit for both X and Z cases. The output error rate of the Z-measurement does not increase very rapidly, however. This behavior qualitatively may be understood as the code maintaining a $\sim 4$ distance correction for X-like errors combined with a lack of \textit{sensitivity} of the logical Z-measure to logical-Z-like errors. Further discussion of site specific \textit{sensitivity} can be found in appendix \ref{Appendix:FuncSites}.

\subsection{Improved decoding with non-uniform edge weights}
\label{decodeComp}
Minimum weight perfect matching (MWPM) is used for decoding in the simulations. Weights for each edge in the decoder graph are assigned according to available information about error rates. We now investigate the decoder performance for the case that error rates for each gate operation are available, \textit{aware} decoding (i.e., Dijkstra algorithm applied to unique edge weights in the decoding graph), in contrast to assuming an average error rate for the gate operations, \textit{naive} (i.e., edge weights using the average error rate for \textit{i}, \textit{h}, \textit{cx} operations). 

An important way that decoding can improve is through resolving ambiguous syndromes for error chains greater than $(d+1)/2$. Random error rate distributions can offer 'tie breaking' information (see for example appendix \ref{Appendix:Decoder}). Improved error rates using \textit{aware} decoding for random distributions for a phenomenological analysis have already been reported \cite{Tiurev_2023}. The prominence of ambiguous cases that can be resolved using this information is likely code dependent. To further probe the question of relative decoder effectiveness of \textit{aware} decoding, we investigate the response of a heavy hex code (HHC) to these two different decoding approaches. The HHC is a hybrid of a surface code and compass code with flag qubits \cite{LowDegree20,Sundaresan_2023}. This combination provides a useful relative view of the effectiveness of aware decoding for these two different codes.

\begin{figure*}[ht!] %h b,t ! 
    \centering
    \subfloat[\label{fig:Fig6:a} \textit{aware} vs. \textit{naive}]{
    	\includegraphics[width=0.3\textwidth, clip,trim= 0mm 0mm 0mm 0mm]{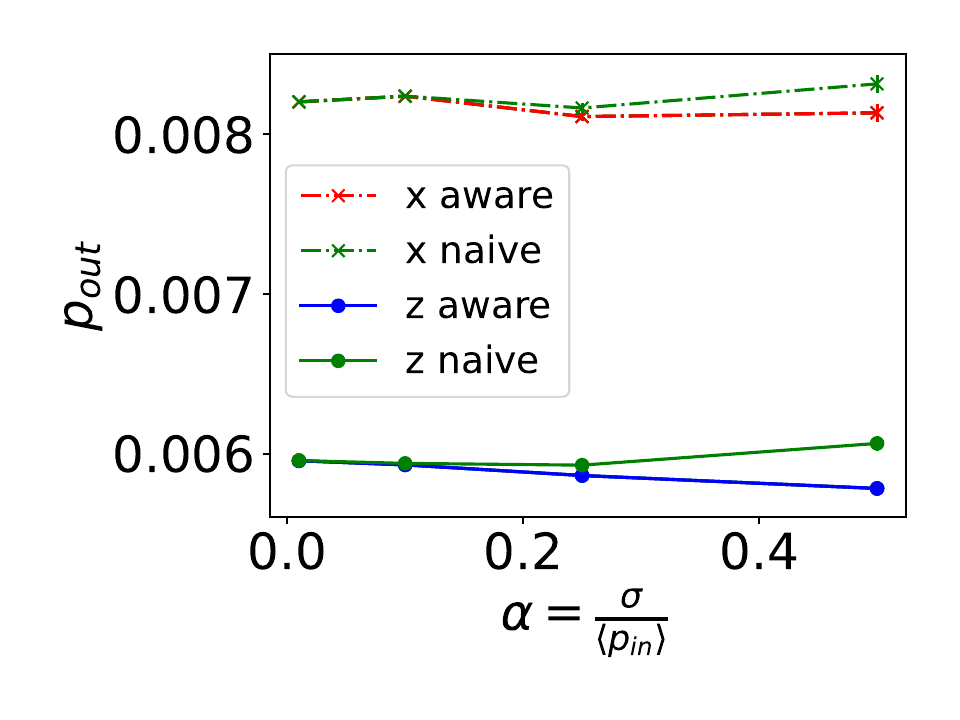}
    }
    \subfloat[\label{fig:Fig6:b} Standard deviation]{
        \includegraphics[width=0.3\textwidth, clip,trim= 0mm 0mm 0mm 0mm]{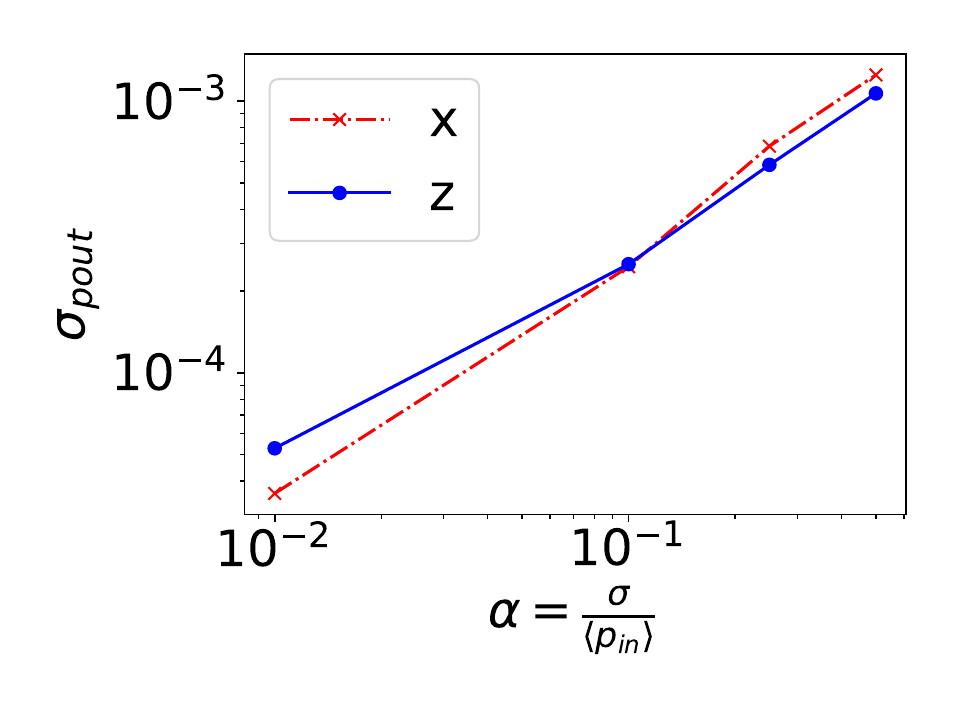}
    }
    \subfloat[\label{fig:Fig6:c}Ratio = $\frac{p_{\textit{naive}}}{p_{\textit{aware}}}$]{
        \includegraphics[width=0.3\textwidth, clip,trim= 0mm 0mm 0mm 0mm]{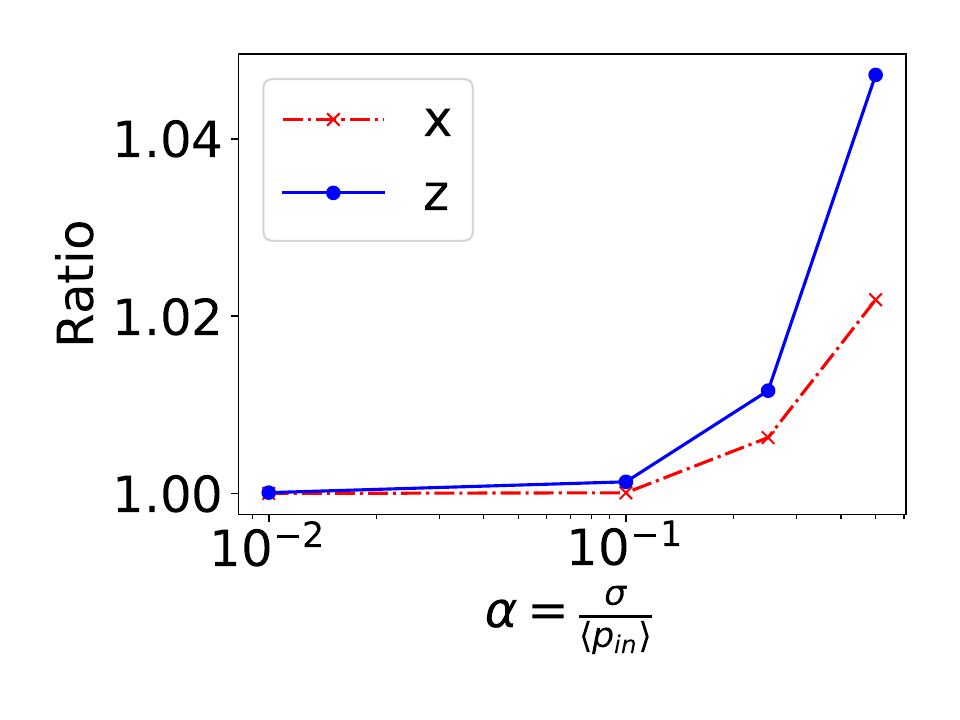}
    }
    \caption{(a) Comparison of \textit{aware}, blue, and \textit{naive}, green, decoding of lattices with i,h and cx operations selected randomly from a normal distribution for multiple distances of the heavy hex code. All other operation error rates are set to a negligible contribution. Results for initialization and measurement in the $|0\rangle$ and $|+\rangle$ basis are labeled as z and x, respectively. The standard deviation of the noise is swept and is set as $\sigma = \alpha \langle p_{in} \rangle$. The cx, i and h operations are set to a constant mean error rate, $\langle p_{in} \rangle = 10^{-3}$. Each point represents 512 iterations of 50k shots. Standard error bars are approximately the size of the symbols. (b) The standard deviation of $\langle p_{out} \rangle$ from the 4 iterations. (c) The ratio between \textit{aware} and \textit{naive} decoders.}
    \label{fig:AwNavDec}
\end{figure*}

Simulations were carried out for \textit{aware} and \textit{naive} and were done for the the $|0\rangle$ (Z) and $|+\rangle$ (X), four different cases, for several distances, Fig. \ref{fig:AwNavDec}. The Z basis corresponds to stronger \textit{sensitivity} to the surface code like checks and the X basis corresponds to compass code like checks with flag qubits. In both X and Z bases, the \textit{aware} decoder reduces $p_{out}$ compared to the \textit{naive} case and the benefit increases with both decreasing $p_{in}$ and increasing $d$, consistent with predictions from phenomenological modeling \cite{Tiurev_2023}. The reduction in $p_{out}$ is, however, relatively weak for x basis cases and is in general relatively small for these small distance codes. Qualitatively weak improvements from decoder improvements for the X basis were reported in experiments on the $d=3$ HHC even when applying maximum likelihood methods. One contribution noted in that work was that the Z check circuit is susceptible to introducing a high rate of Z errors on the data, particularly because of how deflagging is done, which is difficult to improve with better decoding \cite{Sundaresan_2023}. %[would Z error accumulation explain no change in the X basis?].

\section{Discussion}
\label{discussSec}

\subsection{Screening}
During the fabrication process of QPUs, checkpoints can be established where infidelities of qubit operations can be estimated \cite{Zhang_LaserAnneal_22}. Imperfect fabrication processes lead to sets of QPUs each with their own estimated infidelity distributions. \textit{Ranking} of the QPUs within these sets is useful for selection of which QPUs to ultimately deploy. A predictive pre-selection is desirable, all other considerations assumed to be the same.In the context of these findings, a paradigm of \textit{ranking} by average input error rate of a dominant faulty circuit operations (e.g., two qubit gate operations) may therefore be an effective starting strategy in the absence of other supporting indicators like logical qubit error simulation. 

This strategy might be further refined in the presence of a model for the logical error rate dependence on average input error rate, for example, circuit simulations of a particular code and device layout. In such circumstances, more quantitative evaluation of the \textit{sensitivity} to differences in average input error, QPU to QPU, come into consideration including a better understanding of shot to shot standard deviation in the logical error rate.

\subsection{\textit{Sensitivity} to 'bad' location}

The impact of one or a limited number of 'bad' locations is a common question due to many factors ranging from fabrication imperfections to instability in device performance (e.g., spectral diffusion of two level systems \cite{carroll_dynamics_2022}). Here we loosely define a 'bad' location as a faulty location for which the infidelity of the gate operation appears to be an outlier relative to the average error rate.

We have observed, in section \ref{infdist_bad}, that the \textit{sensitivity} of the logical error can be relatively weak to an outlier location until the error rate is appreciably larger than the contribution of the average error distribution. The 'bad' location contribution to the logical error rate is, furthermore, limited as the error rate saturates at roughly the equivalent of the reduction of the code distance by one.  

For a simplified order of magnitude estimation example, we might consider identification of 'bad' two qubit gate locations by assuming a case where the two qubit gate locations are the dominant error rate for the logical error. Then a two qubit location is identifiable as a 'bad' location when its error rate is: $\epsilon_{bad} \sim N_{q}\bar\epsilon$, where $N_{q}$ is the number of qubits in the encoding. In general the average error distribution has many contributions from the different circuit operations. Each of the operation types have their own relative contributions to the logical error rate.  

\subsection{Time dynamic noise}
Error rates are time dependent in realistic devices. There are a number of sources of time dependence \cite{witzel_quantum_2012, muller_towards_2019, de_graaf_two-level_2020,krantz_quantum_2019}. A natural question is what logical error rates to expect in the presence of time dynamic noise. In many cases the time dynamics are correlated, which is out of the scope of this analysis. This work, however, does provide insight about the limit of uncorrelated time fluctuations. In this context, the observation that the average logical error rate is not dependent on higher moments of the distribution provides an indication that the logical error rate will converge to an average based on the average of the input error rates when uncorrelated noise processes are also stationary, while the variance of the logical error rates, in contrast, will depend on higher moments. It is left to future work to establish how close this limit approximates experimental cases that can have time and space correlations.

\subsection{Pseudothresholds and guidance for design}
A device's qubits performance relative to the pseudothreshold of a code is a predominant concern for design, fab and operation of quantum error correction on a QPU. Pseudothresholds are often estimated using simulations with forms of uniform error rates for types of circuit operations (e.g., two qubit operations). How to assess an actual device's non-uniform infidelity distibution relative to a code's pseudothreshold, particularly time varying distributions, without measuring or simulating the specific case (when measurement is not readily available) is a practical problem. This work provides the observation that for uncorrelated noise and surface-like codes, $\langle p_{out} \rangle = F(\bar\epsilon)$ over a relative wide error rate range. The concept of a pseudothreshold therefore is also applicable when framed as an outcome of an average error rate of faulty locations in the error correction circuit. This observation may also be of utility in the context of system design, for which variability in system components are their own source of qubit operation infidelity distributions. System designs therefore will need to assess impact of outliers and variances of their component performances relative to their targets. Analysis based on the average error rates would greatly simplify the challenge of multi-distribution problems, in contrast to analysis of the contributions of multiple distribution each with their own multi-moment parameterizations.

\subsection{Modularity}
Interest has increased recently in noisy connections between devices to form extended modular quantum error corrected patches \cite{ramette_fault-tolerant_2023,auger_fault-tolerance_2017, nickerson_topological_2013}. The impact of a sparse number of high error rate locations (i.e., connection points) on output logical error rates is of central interest in order to provide guidance about design and fabrication tolerances. The logical $\langle p_{out} \rangle$ dependence on 'bad' qubit sites placed along a logical operator, in this work, is of tangential relevance as it highlights that if the intersecting 'seam' between modules is placed along a logical operator, it may unnecessarily exaggerate the deleterious impact of the 'seam' compared to staggering the intersections in a less spatially correlated mode (i.e., sparse random). 

\section{Conclusion}
We have studied independent non-identical noise distributions at circuit level for proxy cases of the surface and compass code families. The circuits are mapped to a heavy hex layout and for this work it is shown how to map a rotated subsystem surface code (RSSC) onto the heavy hex lattice. These cases are of general interest to understand surface and compass code trends, while the  particular cases are of direct interest to the application of superconducting qubit quantum error correction \cite{Sundaresan_2023}.

A central result is that the $\langle p_{out} \rangle \approx F(\langle p_{in} \rangle)$ for the distributions studied in this work, where $\langle p_{in} \rangle$ is the average error rate of one or a combination of dominant faulty circuit operations. Notably, $\langle p_{out} \rangle$ is not a strong function of higher moments of the error distribution over the range of parameters studied. The standard deviation of $p_{out}$, $\sigma_{p-out}$, however, does depend on higher moments of the distribution. The independence of $\sigma_{p-out}$ is consistent with a phenomenological model of the repetition code which, for a three qubit example is: $\langle p_{out}^{rep} \rangle = 3\bar\epsilon^2 - 2\bar\epsilon^3$ (see appendix \ref{appendix:repModel}).

The effect of outliers is examined with simulations of the \textit{sensitivity} to varying error on a single or few 'bad' sites. The simulations highlight that $\epsilon_{bad} \sim \mathcal{O}(N_q\bar\epsilon)$ begin to appreciably change the logical error rate. The logical error increase, furthermore, is limited and saturates. The saturation qualitatively behaves as if the code distance is reduced by the introduction of the 'bad' qubit, $\epsilon_{bad} \gg \mathcal{O}(N_q\bar\epsilon)$ .  

For reference, we discuss a proxy physical example of non-identical error rate, decoherence from energy relaxation which, for example, produces a time dynamic distribution of error rates in superconducting qubits. A normal distribution loosely fits the histogram frequency of $T_1$ measured across devices. We use this example to highlight the importance of the harmonic mean of coherence times as being a relevant measure for estimating $\langle p_{out} \rangle$. A number of studies in the literature emphasize normally distributed infidelity distributions, which may overlook the importance of the role of higher moments in $p_{in}$ (e.g., $\sigma$). Error rates are often reciprocally dependent on coherence times, for example, which are determined by common  intrinsic (e.g., fab defects) and extrinsic (e.g., electronics noise) distributions of noise. 

Tantalizingly, the non-identical distributions of error rates offer an opportunity in decoding approaches like minimum weight perfect matching. Decoding of syndromes can be hampered by ambiguities for error chains of $(d+1)/2$ or greater with the same syndrome. Non-uniform error rates introduce additional information, \textit{aware}, that might improve decoding relative to \textit{naive}. As with previous investigators, we observe an improvement in decoding with \textit{aware} information relative to \textit{naive}. For normally distributed error rates across the probe operations (i.e., $i, h$ and $cx$ operations), the improvement increases with increasing $\sigma$ of the error rate distribution. It is a possibly interesting future direction to assess how much of a decoding utility the non-uniformity can be.  

In summary, we have numerically studied non-identical, independent noise distributions for surface and compass-like codes at a circuit level. We observe several trends that provide heuristic guidance regarding the role of the average in contrast with higher moments of error rate distributions play on the logical error. We also probe the \textit{sensitivity} of logical error rates to one or a few outlier error rate device locations, which provides additional insight about the role of 'bad' locations and identification of error rates at these locations that cross-over to being non-negligible relative to the weight of the rest of the error contributions in the circuit. These observation provide practical insights into areas ranging from screening, modularity and system design.

\section{Acknowledgements}
We thank Ted Yoder for improvements to the circuit shown in Fig.~\ref{fig:gauge-circuit:b}. We thank Kenny Tran for access to computing resources for the numerical simulations. We acknowledge insightful discussions with A. Corcoles, B. Brown, L. Govia, M. Takita, D. McKay, J. Tersoff and T. Yoder. Parts of this research was sponsored by the Army Research Office accomplished under Grant Number W911NF-21-1-0002 and by IARPA under LogiQ (contract W911NF-16-1-0114). The views and conclusions contained in this document are those of the authors and should not be interpreted as representing the official policies, either expressed or implied, of the Army Research Office or the U.S. Government. The U.S. Government is authorized to reproduce and distribute reprints for Government purposes notwithstanding any copyright notation herein.

\section{References}
% \bibliography{MyLibrary.bib}
\bibliography{ZoteroLib.bib,Library.bib,ACpaper.bib}

\newpage
\newpage

\appendix

\section{$T_{1}$ coherence example}
\label{appendix:infdist_T1}
\subsection{Averaged $T_{1}$ distribution}
Energy relaxation is a leading source of decoherence in superconducting qubits. A spread of relaxation times, $T_1$, are observed across many qubit devices at present. The single and two qubit gate infidelities depend on these coherence times leading to a source of non-uniform infidelity distribution across a device. The energy relaxation rates, furthermore, fluctuate in time over a wide range of time scales \cite{klimov_fluctuations_2018,muller_towards_2019,paladino_mathbsf1mathbsfitf_2014}.

An example distribution for a 20 qubit device is shown in figure \ref{fig:Fig1} (a) \cite{carroll_dynamics_2022}. The distribution is representative of several common features of $T_1$ distributions observed in devices. We note that the histogram and 'normal'-like distribution is, however, not necessarily an absolutely complete representation of the actual $T_1$ distribution. That is the distribution (1) may include some cases that are non-exponential decays due to strong couplings to defects; (2) may miss some fluctuations due to limited measurement time resolution of a non-white power spectral distribution; and (3) may miss some cases in the low coherence time tail of the distribution because the $T_1$ was immeasurably short. 

In this work we primarily focus on the normal-like part of the distribution as this is sufficient to highlight some of the key implications of independent but non-identically distributions (i.n.d.) for quantum error correction. We note that there are a variety of sources of non-uniform infidelity distribution in devices \cite{krantz_quantum_2019,paladino_mathbsf1mathbsfitf_2014}. The energy relaxation distribution is a proxy for the more general problem of non-uniform, fluctuating infidelity distributions.

\subsection{How normal are the measurable $T_1$ distributions?}
\begin{figure}[]%[htbp]
    \centering
	\includegraphics[width = 8.5 cm, clip,trim= 0mm 40mm 0mm 30mm ]{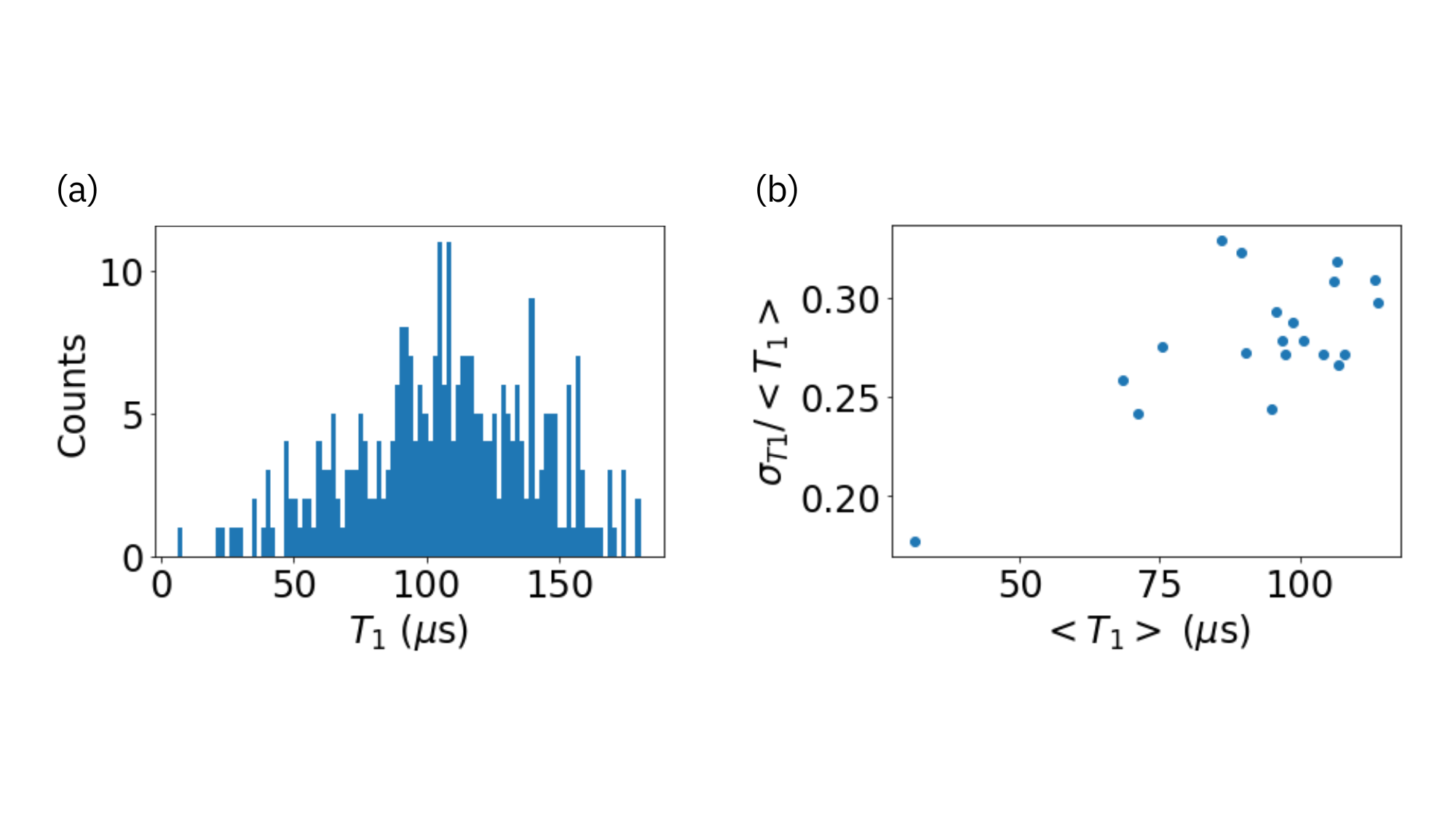}
    \caption{(a) histogram of $T_1$s measured on qubit labeled 0 of the 20 qubits described in Carroll et al. \cite{carroll_dynamics_2022} and in Fig. \ref{fig:Fig1} in the main text. (b) The ratio $\sigma_{T1} / \langle T_1 \rangle$ plotted according to its corresponding the qubit's $ \langle T_1 \rangle$ for the device.}
	\label{fig:histo1Q}
\end{figure}

The $T_1$ times for all qubits were collected for a period over 9 months, a set of 6140 values, shown in the main text, Fig. \ref{fig:Fig1}. A representative histogram of a $T_1$ measured daily for just one qubit is shown in Fig. \ref{fig:histo1Q}. A Kolmogorov-Smirnov null hypothesis test was done for each qubit, for which the null hypothesis is that the $T_1$ distribution is indistinguishable from a normal distribution. The test statistic was 0 and the p-value was greater than 0.999 indicating the normal distribution was indistinguishable from the measured $T_1$ distribution to the limits of accuracy of this statistical measure. We caution, as noted above, that the $T_1$ distribution has some bias through neglect of outlier instances when the qubit's $T_1$ is so short that it is immeasurable.  

The standard deviation of the $T_1$ distribution for each qubit is shown in Fig. \ref{fig:histo1Q} (b). Most qubits show standard deviations of $\sim 0.25-0.3 \langle T_1 \rangle$. The details of the $T_1$ measurement are described in a previous publication \cite{carroll_dynamics_2022}.

\section{Mean of distribution dependent on reciprocal of normally distributed random variable}
\label{appendix:RecipOfNormDist}
Many error rates depend on the reciprocal of a parameter, for example when the error rate is dominated by coherence time it can be approximated as $\epsilon \propto 1/\tau$, where $\epsilon$ is the error rate and $\tau$ is a coherence time. 

In this appendix we are concerned with the error rate mean when the {\textit{'reciprocal parameter'}} is a normally distributed random variable.  We will show that for a normally distributed {\textit{'reciprocal parameter'}} (e.g., $T_1$) that the new random variable's mean is no longer constant rather it depends on $\sigma$, the standard deviation of the {\textit{'reciprocal parameter'}}.

We start by defining a {\textit{'reciprocal parameter'}} random variable {\bf{$T_1$}} that is normally distributed:
\begin{equation}
\begin{aligned}
P({\bf{T_1}}=T_1) = \frac{1}{\sqrt{2\pi\sigma}}\exp{\frac{-(T_1 - \mu)^2}{2\sigma^2}}
\end{aligned}
\end{equation}
\label{eqn:normalT1}
where $\mu$ is $\langle T_1 \rangle$ and $\sigma$ is the standard deviation of the distribution. 

We now consider a new random variable, the error rate, formed with the {\textit{'reciprocal parameter'}}, ${\bf{e}} = t_{gate}/{\bf{|T_1|}}$. The probability of a qubit site having an error rate of $P({\bf{e}}=e(T_{1}))$ is therefore $P({\bf{T_1}}=T_{1})$. 

\begin{figure}[] %h b,t !
    \centering
    \subfloat[\label{fig:Fig15:a}Comparison of distributions]{
    	\includegraphics[width=0.22\textwidth, clip,trim= 0mm 0mm 0mm 0mm]{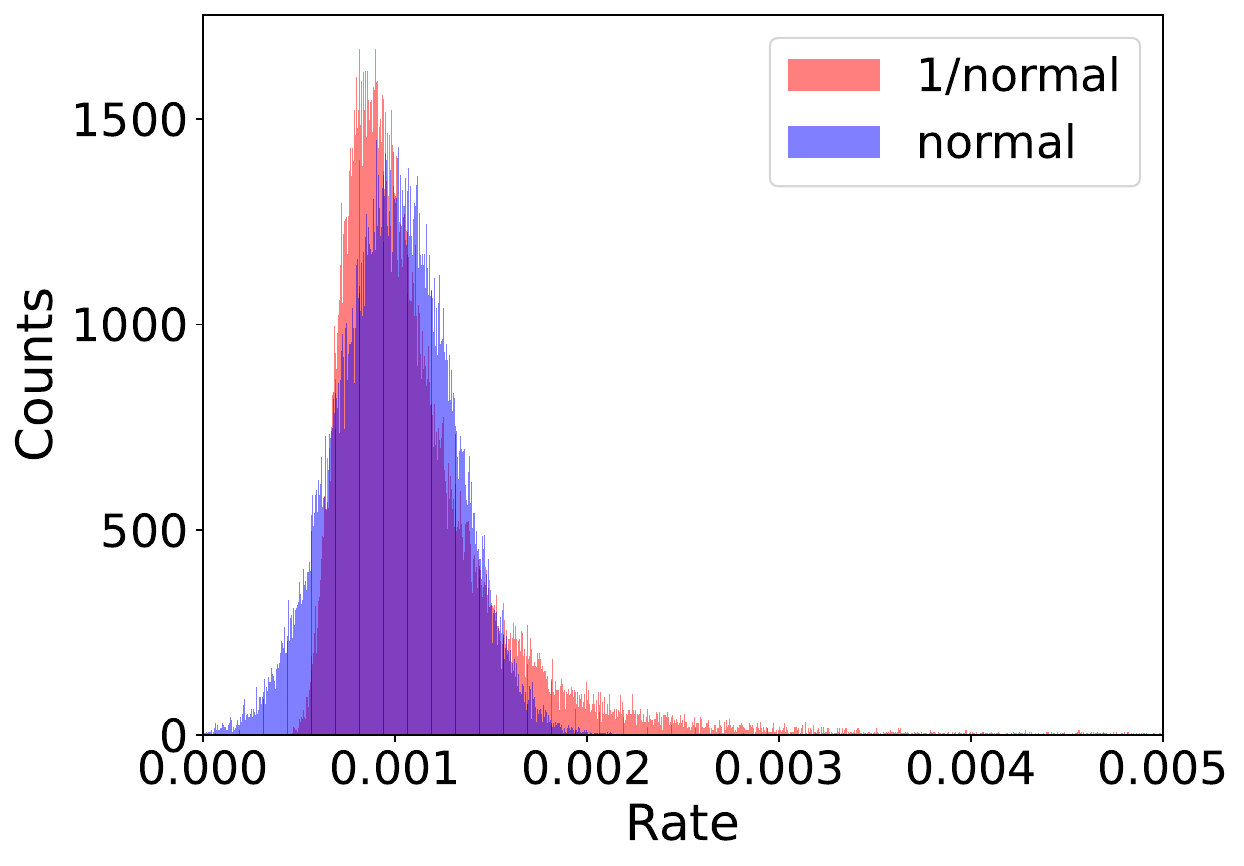}
    % 	\caption{histogram for $\alpha=0.3$}
    }
    \subfloat[\label{fig:Fig15:b} Mean shift by $\alpha$]{
        \includegraphics[width=0.22\textwidth, clip,trim= 0mm 3mm 0mm 0mm ]{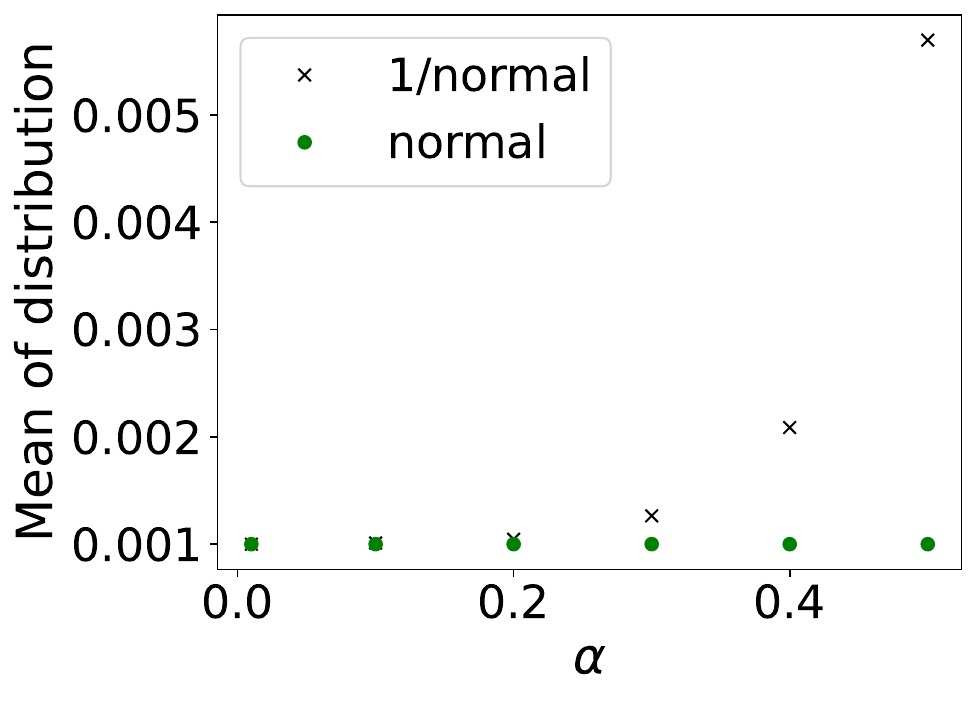}
        % \caption{mean dependence on $\alpha$}
    }
    \caption{(a) Histogram of error rates produced with either a normal distribution ($\mu_0 = 10^{-3}$ and $\sigma = 0.3 \mu$) or from a scaled reciprocal normal distribution with the same $\sigma$. That is, $f =$ normal$(\mu_1,\sigma)$ and the error is $1/f$, where $\mu_1=1/\mu_0$. (b) The resulting mean error of the distribution as a function of the standard deviation, $\sigma = \alpha \mu$.} 
    \label{fig:PoutMeans}
\end{figure}

The mean of the error rate distribution is:
\begin{equation}
\begin{aligned}
\langle e \rangle = \int_{-\infty}^{\infty} \frac{\tau}{T_1}P(T_1) \,dP(T_1)\ \\
 = \int_{-\infty}^{\infty} \frac{\tau}{T_1}\frac{1}{\sqrt{2\pi\sigma}}e^{\frac{-(T_1 - \mu)^2}{2\sigma^2}}d(T_1)
\end{aligned}
\label{eqn:intAvg}
\end{equation}
To see analytically that the mean will depend on $\sigma$ we evaluate the integral in the approximation of very small $\sigma$ for which we may Taylor expand $\frac{\tau}{T_1}$ around $\mu = \langle T_1 \rangle$. This simplification allows integration over all values but is implicitly inaccurate for $T_1$ approaching 0, however, it qualitatively shows the trend for the largest weight of the distribution. The average error in this simplistic approximation is:

\begin{equation}
\begin{aligned}
\langle e \rangle \approx \frac{\tau}{\sqrt{2\pi}} \int_{-\infty}^{\infty}\left[\frac{1}{\mu} - \frac{(T_1-\mu)}{\mu^2}+ \frac{(T_1-\mu)^2}{\mu^3} + \dots \right] \\ \times e^{\frac{-(T_1 - \mu)^2}{2\sigma^2}}\frac{dT_1}{\sigma} 
\end{aligned}
\end{equation}
Using well established identities for the integration of the product of the Gaussian function and polynomials, for example:

\begin{equation}
\begin{aligned}
\frac{1}{\sqrt{2\pi}} \int_{-\infty}^{\infty}x^2 e^{-(ax)^2}dx= \frac{1}{2a}\left(\frac{\pi}{a}\right)^\frac{1}{2} 
\end{aligned}
\end{equation}
where $a=\frac{1}{2}$, $x = \frac{(T_1 - \mu)}{\sigma}$ and $dx = \frac{dT_1}{\sigma}$. To 2nd order, dropping higher terms, the average of the error becomes:

\begin{equation}
\begin{aligned}
\langle e \rangle \approx \tau\left(\frac{1}{\mu} +  \frac{\sigma^2}{8 \mu^3}\right) 
\end{aligned}
\label{eqn:intAvgApprox}
\end{equation}
for which we see that $\langle e \rangle$ increases as a quadratic function of $\sigma$. Odd polynomials do not contribute to the integration over the even Gaussian interval and better approximation would include higher even order terms. Overall the 2nd order approximation grossly underestimates the quantitative dependence of $\langle e \rangle$ on $\sigma$ because the approximation neglects contributions from the $T_1$ values closer to 0 that can be very large but qualitatively confirms that an anticipated increase in average error as $\sigma_{T1}$ increases. For reference we numerically evaluate $e = \frac{\tau}{T_1}$ truncated for $T_1 > 1  ~\mu$s and $\tau=100 ~ns$ in Fig. \ref{fig:PoutMeans} and show its dependence on $\sigma$.

\section{Numerical threshold estimates for the RSSC on the heavy-hexagon lattice} \label{appendix:RSSC}

\begin{figure*}[]
\centering
 \subfloat[Right-pointing gauge operators \label{scheduled-gauge-circuit:a}]{
    %\begin{tikzpicture}  % keep this source code
    %\begin{yquant}
    %qubit {$|\textrm{up}\rangle$} q;
    %qubit {$|0\rangle$} a;
    %qubit {$|\textrm{right}\rangle$} q[+1];
    %qubit {$|0\rangle$} a[+1];
    %qubit {$|\textrm{down}\rangle$} q[+1];
    %box {I} q[0];
    %box {I} q[1];
    %cnot a[1] | q[2];
    %cnot a[0] | q[0];
    %cnot q[1] | a[1];
    %box {I} q[2];
    %box {I} q[0];
    %cnot a[0] | q[1];
    %box {I} a[1];
    %box {I} q[2];
    %box {I} q[0];
    %measure a[0];
    %cnot q[1] | a[1];
    %box {I} q[2];
    %box {I} q[0];
    %box {I} q[1];
    %cnot a[1] | q[2];
    %\end{yquant}
    %\end{tikzpicture}
    \includegraphics[width=0.25\textwidth]{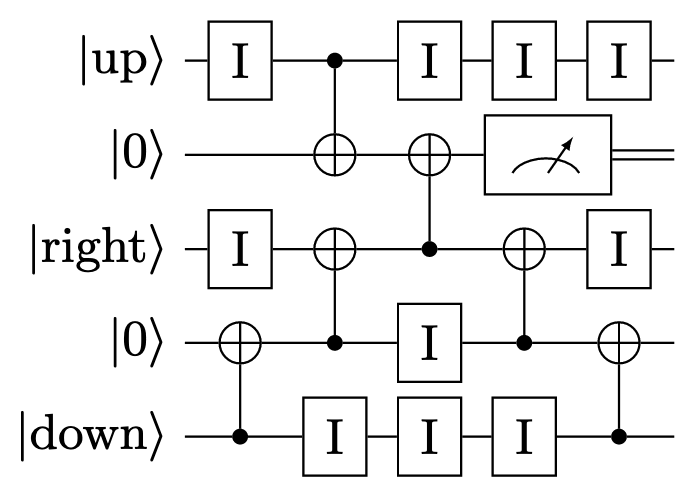}
    }
\qquad
\subfloat[Left-pointing gauge operators \label{scheduled-gauge-circuit:b}]{
    %\begin{tikzpicture}  % keep this source code
    %\begin{yquant}
    %qubit {$|\textrm{up}\rangle$} q;
    %qubit {$|0\rangle$} a;
    %qubit {$|\textrm{left}\rangle$} q[+1];
    %qubit {$|0\rangle$} a[+1];
    %qubit {$|\textrm{down}\rangle$} q[+1];
    %box {I} q[2];
    %box {I} q[1];
    %cnot a[0] | q[0];
    %cnot a[1] | q[2];
    %cnot q[1] | a[0];
    %box {I} q[0];
    %box {I} q[2];
    %cnot a[1] | q[1];
    %box {I} a[0];
    %box {I} q[0];
    %box {I} q[2];
    %measure a[1];
    %cnot q[1] | a[0];
    %box {I} q[0];
    %box {I} q[2];
    %box {I} q[1];
    %cnot a[0] | q[0];
    %\end{yquant}
    %\end{tikzpicture}
    \includegraphics[width=0.25\textwidth]{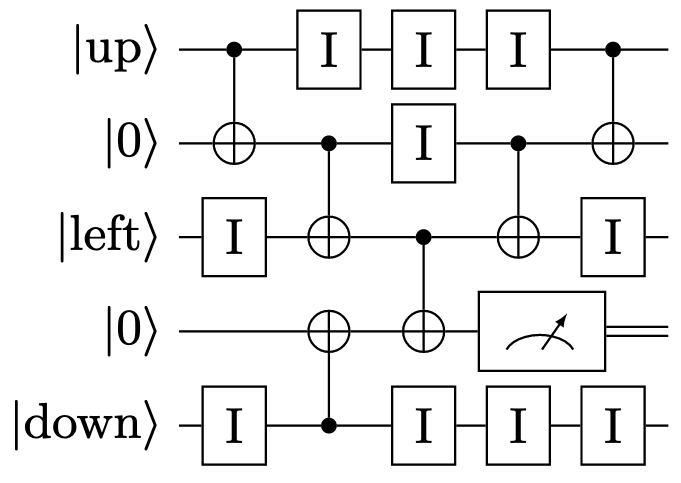}
}
\qquad
\subfloat[Boundary gauge operators \label{scheduled-gauge-circuit:c}]{
    %\begin{tikzpicture}  % keep this source code
    %\begin{yquant}
    %qubit {$|\textrm{up}\rangle$} q;
    %qubit {$|0\rangle$} a;
    %qubit {$|0\rangle$} a[+1];
    %qubit {$|0\rangle$} a[+1];
    %qubit {$|\textrm{down}\rangle$} q[+1];
    %box {I} q[0];
    %box {I} q[1];
    %cnot a[2] | q[1];
    %box {I} q[0];
    %cnot a[0] | q[0];
    %cnot a[1] | a[2];
    %box {I} q[1];
    %box {I} q[0];
    %cnot a[1] | a[0];
    %cnot a[2] | q[1];
    %cnot a[0] | q[0];
    %measure a[1];
    %box {I} q[1];
    %\end{yquant}
    %\end{tikzpicture}
    \includegraphics[width=0.25\textwidth]{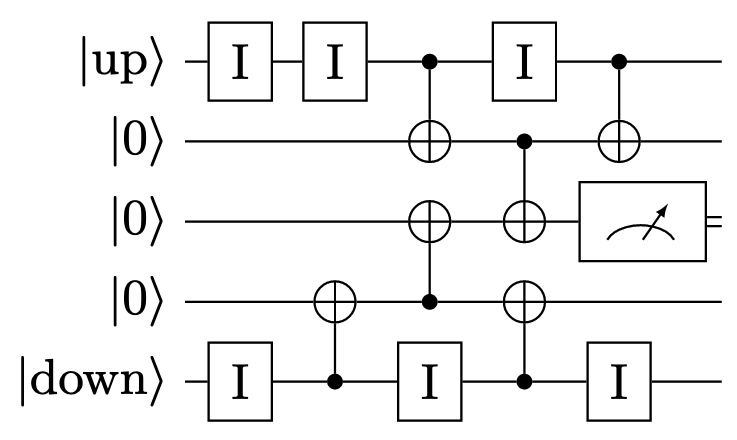}
}
\caption{{\bf Scheduled circuits for measuring Z-type gauge operators}. (a) The second auxiliary qubit is reset in the prior circuit, so we need not account for idle time during that reset operation. (b) Similarly, the first auxiliary qubit is reset within the prior. (c) On the boundary there are three auxiliary qubits and two data qubits.}
\label{fig:scheduled-gauge-circuit}
\end{figure*}

\begin{figure}[]%[htbp]
    \centering
	\includegraphics[width=0.3\textwidth]{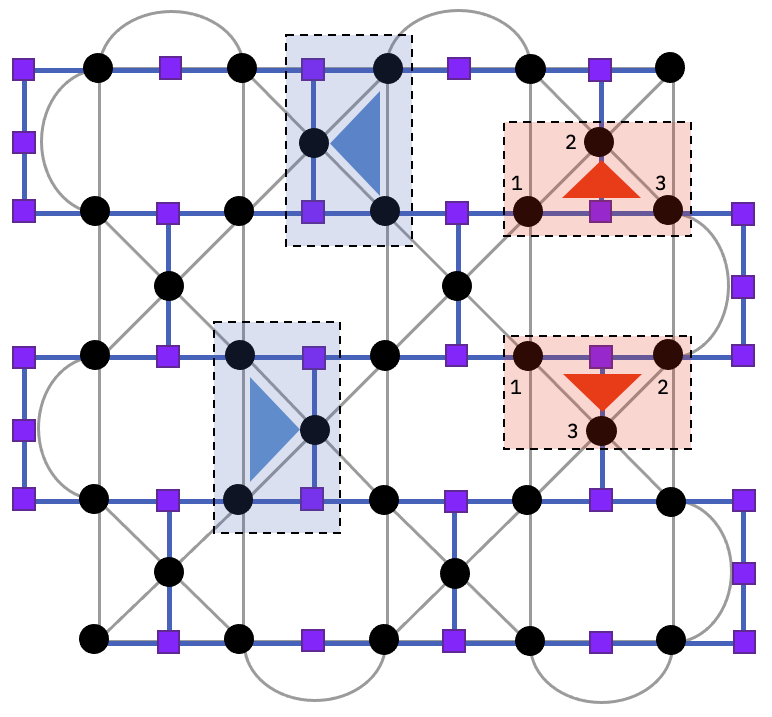}
    \caption{\textbf{Measuring RSSC checks on the heavy hexagon lattice.} Each gauge operator type (X, Z) has two distinct orientations (up/down, left/right). X gauge operators (red) can be measured in parallel using the circuit in Fig.~\ref{fig:gauge-circuit:a}. The numbers label the time step for each CNOT gate. Z gauge operators (blue) with the same orientation can be measured in parallel with the circuit in Fig.~\ref{fig:gauge-circuit:b} since those operators involve disjoint data and auxiliary qubits.}
	\label{fig:layout}
\end{figure}

The RSSC syndrome measurement circuit is a sequence of X and Z gauge operator measurement cycles. The X measurements are scheduled in 3 parallel CNOT layers using time steps shown in Fig.~\ref{fig:layout}. The Z measurements occur in two stages where we measure all of the left-pointing triangles followed by all of the right-pointing triangles. In each stage, the circuits act on disjoint sets of qubits, so we can schedule them independently according to Fig.~\ref{fig:scheduled-gauge-circuit}.

To evaluate the logical error probability, we construct quantum memory circuits that prepare, store, and measure logical $|0\rangle$ or $|+\rangle$ states. These circuits prepare all of the data qubits in either $|0^n\rangle$ or $|+^n\rangle$, respectively, measure X and Z gauge operators in some sequence, and measure all of the data qubits in the $Z$ or $X$ basis, respectively. We consider sequences of X and Z gauge measurements represented by the string $(Z^sX^s)^t$ for positive integers $s$ and $t$. This string means we apply the Z measurements $s$ times followed by the X measurements $s$ times, and the whole sequence is repeated $t$ times. We set the total number of ZX measurements to $st=12$ and iterate over $s=1, 2, 3, 4$. As in \cite{higgott_subsystem_PhysRevX}, we find that $s=2$ optimizes the logical error rate. In this case, roughly half the syndrome bits are computed from the product of eigenvalues of a pair of gauge operators, while the other half is given directly by the eigenvalues of the gauge operators. The matching subroutine in the decoder is implemented using PyMatching \cite{higgott2020pymatching}.

We simulate these circuits using a standard Monte-Carlo simulation wherein we sample from a collection of faulty circuits. Faulty gates are modeled as ideal gates followed by a Pauli channel that applies with probability $p$ a uniformly random non-identity Pauli error on all output qubits. Each type of operation and idle qubit can fail. Faulty preparations and measurements flip their outputs with probability $p$. The simulation is considered a failure if the logical measurement outcome is incorrect. The logical qubit we store in the memory is exposed to a logical Pauli channel with parameters $p_x$, $p_y$, and $p_z$. When we prepare and measure in the $Z$ basis, the simulation provides an estimate of the logical Pauli error probability $p_x+p_y$, and when we measure in the $X$ basis we estimate $p_z+p_y$.

The logical $X$ and $Z$ operators have minimum weight $d$ and are related by a transversal Hadamard gate followed by a reflection about the diagonal, so they we expect them to have comparable logical error probabilities $p_X=O(p_Z)$ that are suppressed to the same order in the code distance. The logical $Y$ operators of the code also have minimum weight $d$, but there is only one coset representative with this weight, whereas there are $O(2^d)$ coset representatives of logical $X$ weight $d$. Therefore, we expect logical Y errors to be suppressed. For this reason, we choose to (over)estimate the total logical error rate as the total $p_x+p_z+2p_y$ of the estimates from the $Z$ and $X$ basis circuits.

\begin{figure*}[]%[htbp]
    \centering
	\includegraphics[width=0.7\textwidth]{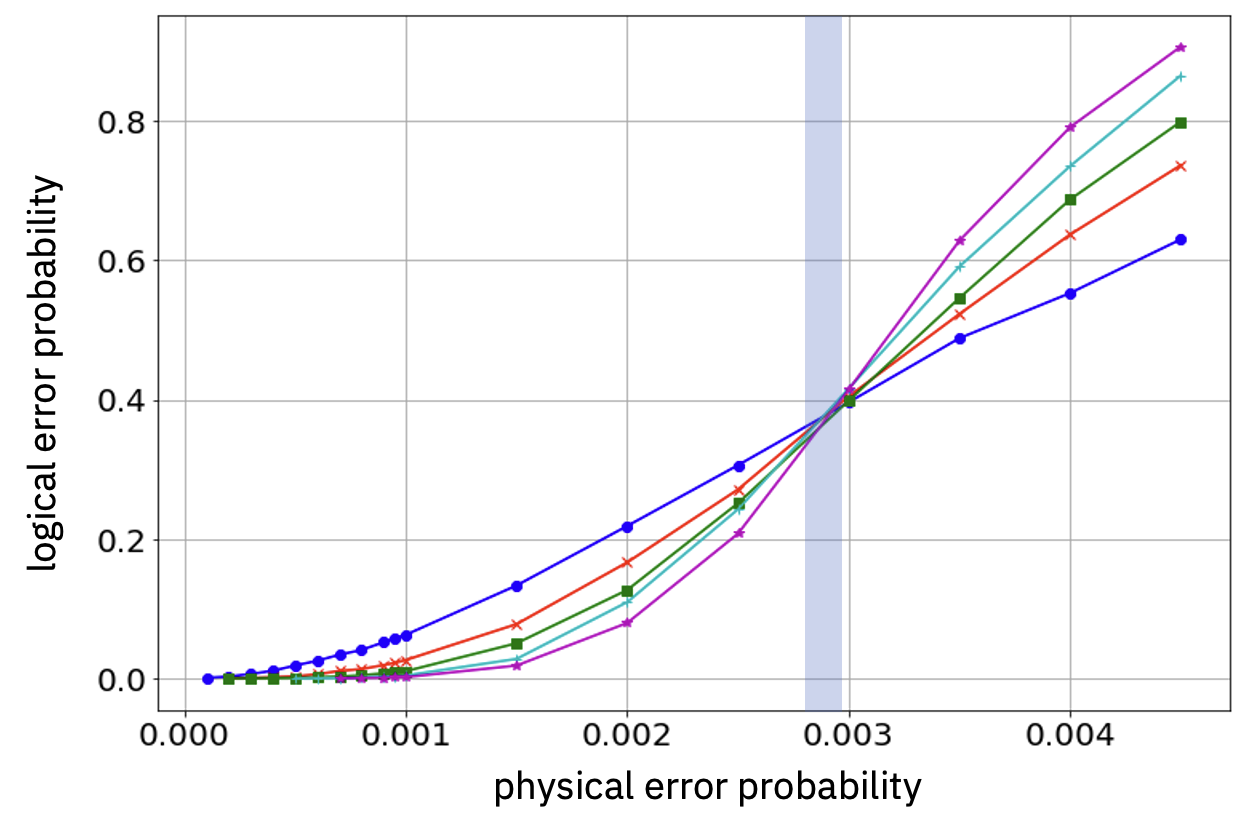}
    \caption{\textbf{Threshold estimate for RSSC on heavy-hexagon lattice.} The total logical error rate $p_x+p_z+2p_y$ is plotted versus the physical error rate for codes with distances $d=3, 5, 7, 9, 11$ (circle, cross, square, plus, star). Data points with the same code distance are connected by line segments. The syndrome measurement schedule $(ZZXX)^6$ corresponds to $2$ gauge measurement repetitions and $12$ total rounds of syndrome measurements. This schedule produces the highest threshold (approximately $0.3\%$) of all schedules we considered. We take 10,000 samples per point for physical error rates $0.0005 \leq p\leq 0.0045$ and 100,000 samples per point for physical error rates $0.0001\leq p < 0.0005$.}
	\label{fig:threshold-hex-optimal-s}
\end{figure*}

\begin{figure*}[]%[htbp]
    \centering
	\includegraphics[width=0.7\textwidth]{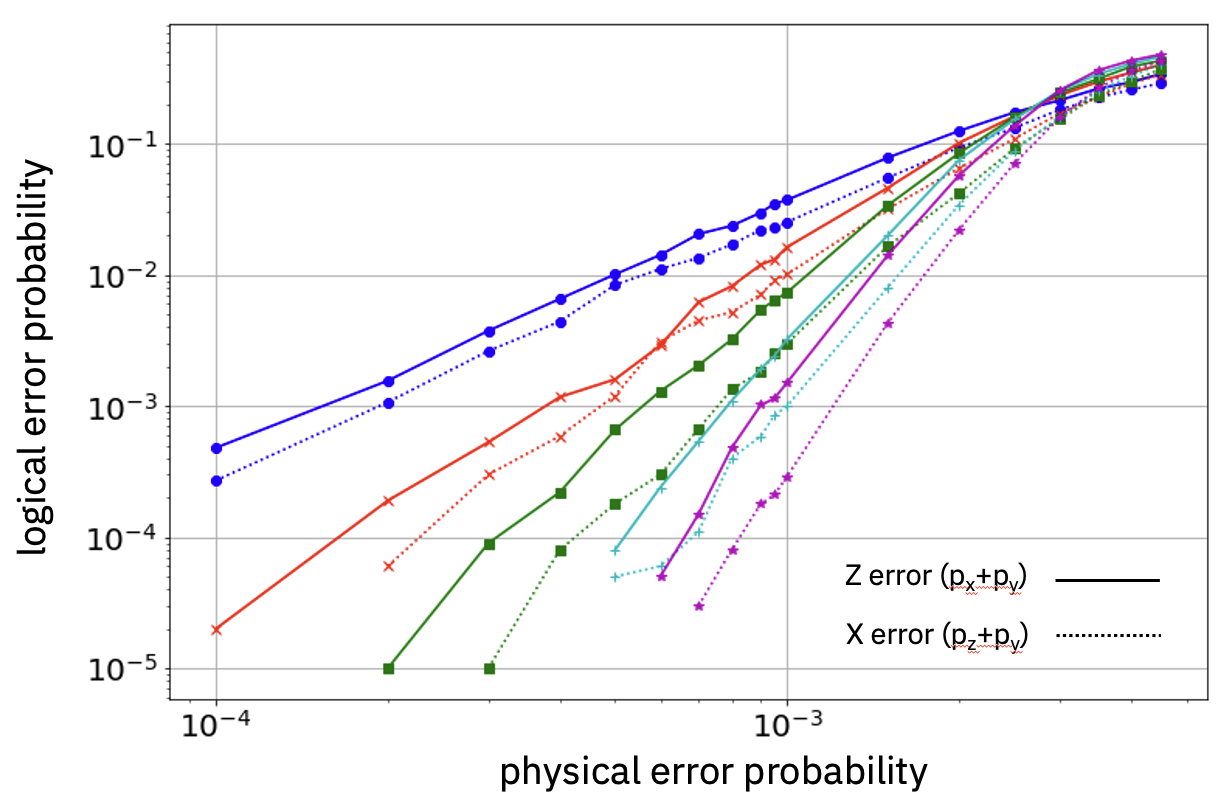}
    \caption{\textbf{Logical error rates for RSSC on heavy-hexagon lattice.} The bit-flip (dotted line) and phase-flip (dashed line) logical error rates are plotted versus the physical error rate for codes with distances $d=3, 5, 7, 9, 11$ (circle, cross, square, plus, star). Data points with the same code distance are connected by line segments. The syndrome measurement schedule $(ZZXX)^6$ corresponds to $2$ gauge measurement repetitions and $12$ total rounds of syndrome measurements. We take 10,000 samples per point for physical error rates $0.0005 \leq p\leq 0.0045$ and 100,000 samples per point for physical error rates $0.0001\leq p < 0.0005$.}
	\label{fig:logical-error-hex-optimal-s}
\end{figure*}

Simulation results are shown in Fig.~\ref{fig:threshold-hex-optimal-s} and Fig.~\ref{fig:logical-error-hex-optimal-s}. These results suggest a threshold of nearly $0.3\%$ for the RSSC on the heavy-hexagon lattice using a circuit-level noise model. A distance-$11$ code is close to break-even at $p=10^{-3}$ and its logical error rate rapidly decreases to below $10^{-4}$ as $p$ approaches $6\times 10^{-4}$.

\begin{figure*}[]%[htbp]
    \centering
	\includegraphics[width=0.7\textwidth]{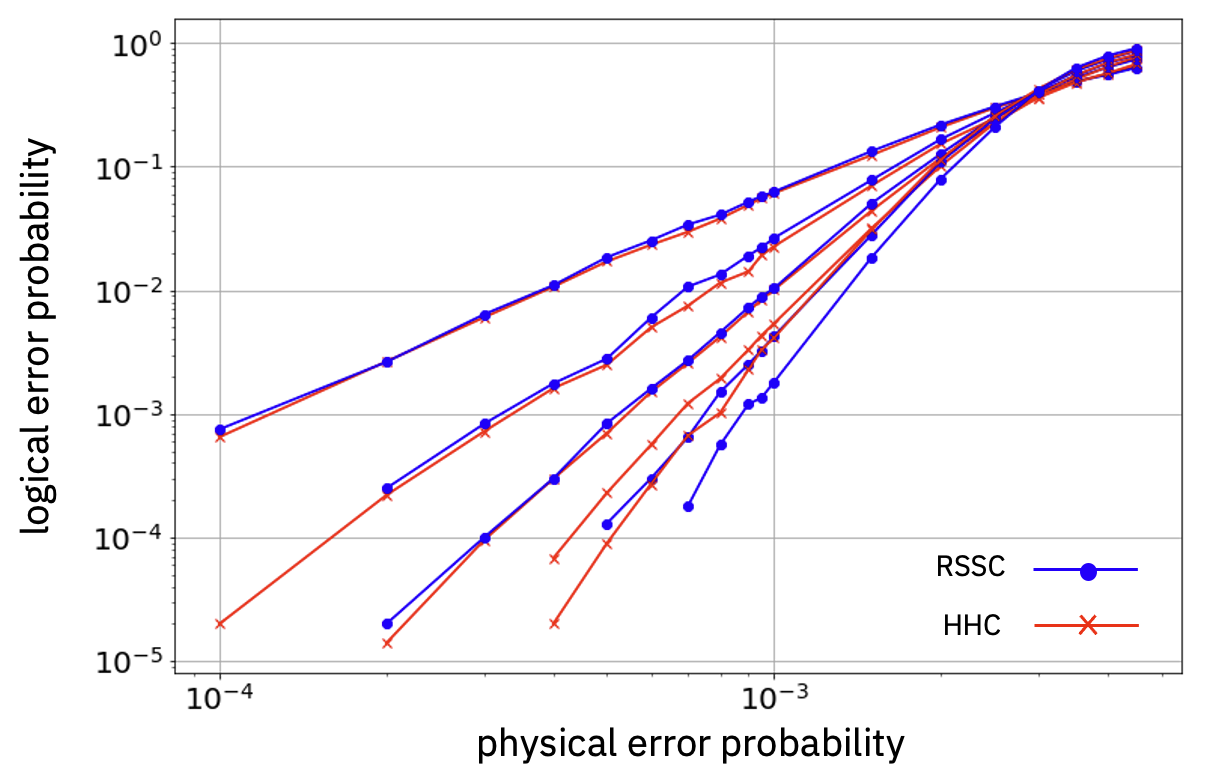}
    \caption{\textbf{Logical error probabilities for RSSC and HHC on heavy-hexagon lattice.} The total logical error rate $p_x+p_z+2p_y$ is plotted versus the physical error rate for codes with distances $d=3, 5, 7, 9, 11$. Data points with the same code distance are connected by line segments. The syndrome measurement schedule $(ZZXX)^6$ corresponds to $2$ gauge measurement repetitions and $12$ total rounds of syndrome measurements. This schedule produces the lowest logical error of all schedules we considered.}
	\label{fig:threshold-compare-s2-bottom}
\end{figure*}

For the purpose of comparison, we carry out simulations of the HHC using the same parameters and syndrome measurement schedules as the RSSC. When we decode the HHC, we choose to use a simple deflagging procedure \cite{Sundaresan_2023} rather than dynamically modifying the edge weights in the decoding graph as in \cite{LowDegree20}. As before, the logical error rate is optimized by choosing $s=2$, so we conclude that both codes' logical error rates are improved by schedule-induced gauge fixing. Figure \ref{fig:threshold-compare-s2-bottom} compares the estimated total logical error rate. We find that the HHC and RSSC have nearly identical total logical error rates for distances $3$, $5$, and $7$, but the RSSC exhibits superior logical error rates at higher distances. This is expected behavior because the RSSC has an asymptotic threshold whereas HHC does not.

\begin{figure*}[]%[h!tbpH]
    \centering
	\includegraphics[height=6.5 cm,width=\linewidth, clip,trim= 0mm 40mm 0mm 40mm]{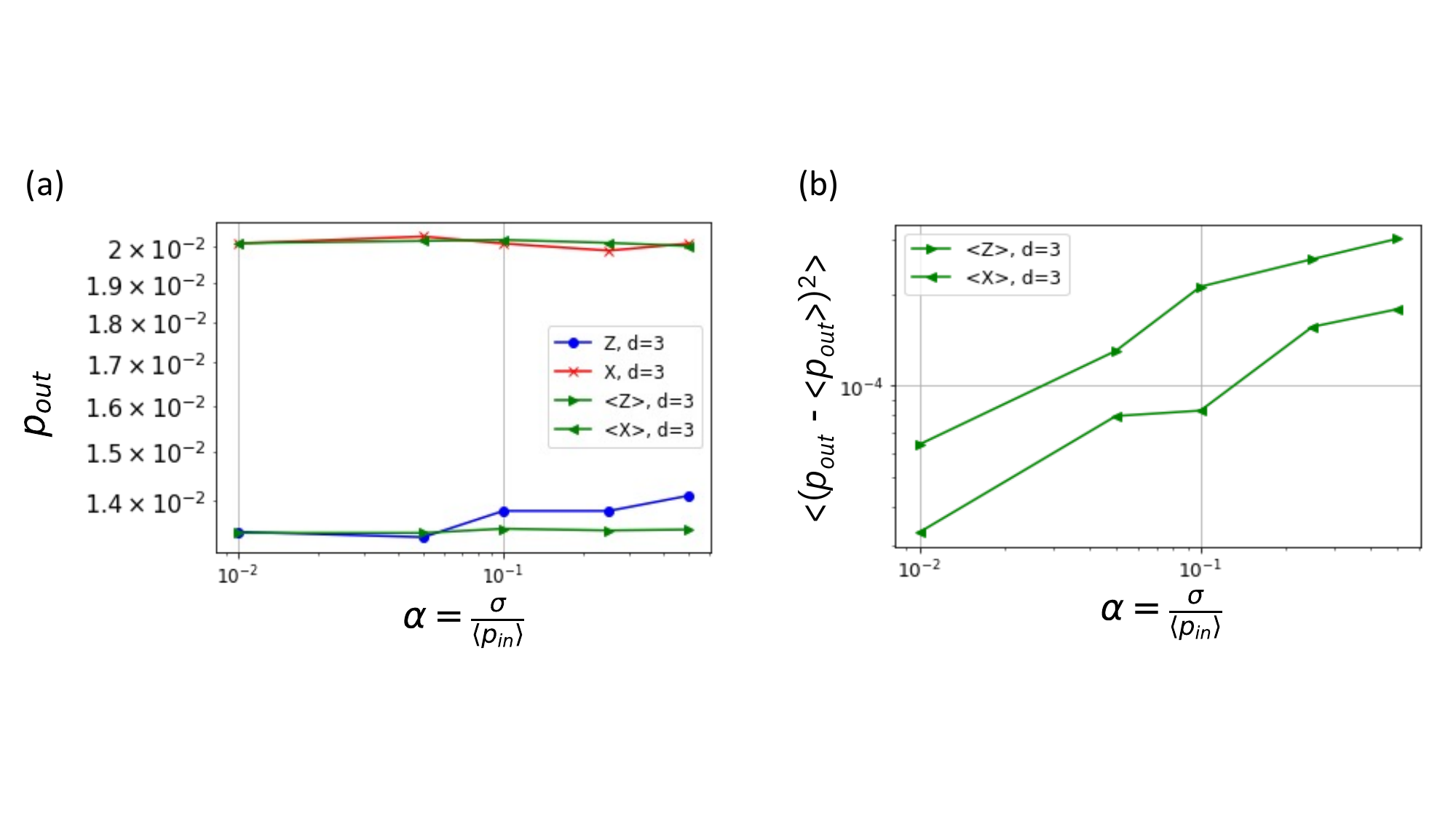}
    \caption{(a) Average logical output dependence on normally distributed error rate, depolarized noise, for the RSSC. The standard deviation of the error rates is parameterized as $\sigma = \alpha \times\langle p_{in} \rangle$. The average is for 64 device instances. The randomly chosen error rate is held constant for each simulation instance. $p_{in} = 10^{-3}$. For reference, one of the instances for each $\alpha$ is overlaid. (b) Standard deviation of the logical output error rates, instance to instance.}
	\label{fig:AvgNormDist}
\end{figure*}

\section{Normal distributed error rate simulations}\label{appendix:normDistSims}

We numerically examine the dependence of the logical output error rate on error rates selected randomly for \textit{h, i} and \textit{cx} at each site for the RSSC. The error rate is drawn from a normal distribution with a standard deviation parameterized as $\sigma = \alpha \langle p_{in} \rangle$ and the error rate is held constant through out the simulated circuit schedule for each instance. The average output error rate converges to the uniform case, Fig. \ref{fig:AvgNormDist} (a). Increasing standard deviation, $\alpha$, in the input error distribution does show a dependence in standard deviation of $\sigma_{out}$, instance to instance, Fig. \ref{fig:AvgNormDist} (b). Qualitatively this is also consistent with what is expected from a phenomenological repetition code, appendix \ref{appendix:repModel}.

\section{Noise model cases of the three qubit code}
\label{appendix:repModel}
\subsection{Case 0: Uniform error for 3 qubit repetition code}
We consider a few simple noise models for a three qubit majority vote code to provide insight about the effect of bias on logical qubit error rate when there is a normally distributed persistent bias selected for each qubit operation. 

We ask what is the error rate, $E$, defined as the probability that the code reports an incorrect output value from the majority vote after a single round of measurements of the three qubits. For a simple case, {\bf{}case 0}, the qubits have the same probability of error, $p$ (e.g., a Bernoulli-like binary trial). The probability of an error can be expressed as $E = 3p^2(1-p) + p^3$.

% and the variance of a sequence of trials with the binary outcome of probability $E$ is $Var[E] = E(1-E)$ (i.e., the variance of a Bernoulli distribution).

\subsection{Case 1: Persistent biased single qubit noise}
We now turn to a second case, {\bf{}case 1}, where we examine how $E$ is effected by adding a random error bias on each qubit site. We substitute unique and persistent error probabilities at each of the $i$ sites, $\epsilon_i$ (e.g., biases on the single qubit operations). The probability of failure for a three qubit example with $\vec{\epsilon} = [\epsilon_1, \epsilon_2, \epsilon_3]$ can then be expressed as,
\begin{equation}
    E(\vec{\epsilon}) = \epsilon_1\epsilon_2(1-\epsilon_0) + \epsilon_0\epsilon_1(1-\epsilon_2) + \epsilon_0\epsilon_2(1-\epsilon_1) + \epsilon_0\epsilon_1\epsilon_2
\end{equation}
simplifying to:
\begin{equation}
    E(\vec{\epsilon}) = \epsilon_0\epsilon_1 + \epsilon_1\epsilon_2 + \epsilon_0\epsilon_2 - 2\epsilon_0\epsilon_1\epsilon_2.
\label{eqn:case1}
\end{equation}
We now define the error rate at the $i^{th}$ qubit as $\epsilon_i = \bar{\epsilon} + \delta_i$ for which $\delta_i$ is the explicit bias error rate on the $i^{th}$ qubit. We consider the case where the $\delta_i$ is persistent shot to shot for a device instance $j$. The unique draw for all $i$ qubits, $\vec{\delta}$, describes an error rate, $E_j$. That is, a particular $j^{th}$ device instance is described by $\vec{\delta}_j$ with an error rate $E_j$. 

We are interested in how $E_j$, the error rate for the $j^{th}$ device instance and $\sigma_{E_j}$ depend on the biased error. More explicitly we have in mind that the probability of drawing $\delta_i$ for a particular site $i$ is described by the function, $f(\delta_i) = \frac{1}{2\pi\sqrt{\alpha\bar{\epsilon}}}e^{-\frac{1}{2}(\frac{\delta_i}{\alpha\bar{\epsilon}})^2}$ (i.e., $\sigma_{\epsilon} = \alpha\bar{\epsilon}$). We now explicitly express the error rate, $E_j$, as a function of the bias error rates for a particular device instance $\vec{\delta}_j$ and in the context of the three qubit example:
\begin{equation}
\begin{aligned}
    E_j(\vec{\delta}_j)=(\bar{\epsilon} + \delta_0)(\bar{\epsilon} + \delta_1) \\ + (\bar{\epsilon} + \delta_0)(\bar{\epsilon} + \delta_2) \\ + (\bar{\epsilon} + \delta_1)(\bar{\epsilon} + \delta_2) \\ -2(\bar{\epsilon} + \delta_0)(\bar{\epsilon} + \delta_1)(\bar{\epsilon} + \delta_2)
\end{aligned}
\end{equation}
and after reorganizing 
\begin{equation}
\begin{aligned}
    E_j(\vec{\delta}_j)=3\bar{\epsilon}^2 + 2(\delta_0 + \delta_1 + \delta_2)\bar{\epsilon} +  (\delta_0\delta_1 + \delta_0\delta_2 + \delta_1\delta_2) - \\ 2(\bar{\epsilon}^3 + (\delta_0 + \delta_1 + \delta_2)\bar{\epsilon}^2 + (\delta_0\delta_1 + \delta_0\delta_2 + \delta_1\delta_2)\bar{\epsilon} + \delta_0\delta_1\delta_2). 
\end{aligned}
\label{eqn:Ej}
\end{equation}

The probability of a particular error rate $E_j$ is now a random variable defined by the sum and product of independent normal distributions of $\epsilon_i$. Different devices will have different biases leading to a set of device instances with error rates of $\{...,E_{j-1},E_j,E_{j+1},\}$. A particular $j^{th}$ instance will have an error rate set by $\bar\epsilon$ and $\vec{\delta}_j$ leading to a sequence of Bernouli 'trials' with average error $E_j$ and $\sigma_{E_j} = \sqrt{E_j(1-E_j)}$.

Through averaging $\langle E_j \rangle = \sum_j^{N_j}{E_j}/N_j$ over the span of the set of $N_j$ device instances  $\{...,E_{j-1},E_j,E_{j+1},\}$ and using the property $\langle XY \rangle = \langle X \rangle \langle Y \rangle$, for independent random variables and $\langle \delta_i \rangle_j \rightarrow 0$ for $N_j \rightarrow \infty$, the error rate averaged over many device instances simplifies eqn. \ref{eqn:Ej} to:
\begin{equation}
     \langle E_j \rangle = 3\bar\epsilon^2 - 2\bar\epsilon^3
\label{eqn:avgEj}
\end{equation}

We also ask how the average variance may depend on $\sigma_{\epsilon}$. The variance is $ \langle \sigma_{E_j}^2 \rangle = \langle (E_j - \langle E_j \rangle)^2 \rangle_j$ averaged over all $j$ instances. Explicitly this becomes:
\begin{equation}
    \begin{aligned}
    \langle \sigma_{Ej}^2 \rangle \sim \langle
    (4\delta_0^2\bar\epsilon^2 + 4\delta_1^2\bar\epsilon^2 + 4\delta_2^2\bar\epsilon^2) \\
    + (\delta_0^2\delta_1^2 + \delta_0^2\delta_2^2 + \delta_1^2\delta_2^2)\\
    - (8\delta_0^2\bar\epsilon^3 + 8\delta_1^2\bar\epsilon^3 +8\delta_2^2\bar\epsilon^3) \\ 
        - (4\delta_0^2\delta_1^2\bar\epsilon+4\delta_0^2\delta_2^2\bar\epsilon + 4\delta_1^2\delta_2^2\bar\epsilon) \\
     + (4\delta_0^2\bar\epsilon^4 + 4\delta_1^2\bar\epsilon^4 +  4\delta_2^2\bar\epsilon^4) \\ 
    +(4\delta_0^2\delta_1^2\bar\epsilon^2 + 4\delta_0^2\delta_2^2\bar\epsilon^2 + 4\delta_1^2\delta_2^2\bar\epsilon^2)\\  +
    4\delta_0^2\delta_1^2\delta_2^2 
    \rangle
    \end{aligned}
\label{eqn:variance1}
\end{equation}
Observing that $\sigma_{\epsilon_i}^2 = \langle (\epsilon_i + \delta_i)^2\rangle - \langle (\epsilon_i + \delta_i) \rangle^2 = \langle \delta_i^2 \rangle$ and keeping only the $4^{th}$ order terms eqn. \ref{eqn:variance1} simplifies to:
\begin{equation}
    \begin{aligned}
    \langle \sigma_{Ej}^2 \rangle \sim 
    4(\sigma_{\epsilon_0}^2\bar\epsilon^2 + \sigma_{\epsilon_1}^2\bar\epsilon^2 + \sigma_{\epsilon_2}^2\bar\epsilon^2)  + \\ (\sigma_{\epsilon_0}^2\sigma_{\epsilon_1}^2 + \sigma_{\epsilon_0}^2\sigma_{\epsilon_2}^2 + \sigma_{\epsilon_1}^2\sigma_{\epsilon_2}^2)
    \propto \langle (\alpha E)E \rangle
    \end{aligned}
\label{eqn:variance}
\end{equation}
where for illustration we parameterize the standard deviation as $\sigma = \alpha \langle E \rangle$ as in the body of the paper. The variance averaged over all instances depends on the standard deviation, $\sigma_{\epsilon}$ and higher moments of $\delta_i$ in contrast to the $\langle E_j \rangle$ that depends only on the input average error rate $\bar\epsilon$.

\section{Output error \textit{sensitivity} to different functional sites}
\label{Appendix:FuncSites}
\begin{figure*}[] %h b,t ! 
    \centering
    \includegraphics[width = \linewidth, height=13 cm, clip,trim= 0mm 0mm 70mm 0mm]{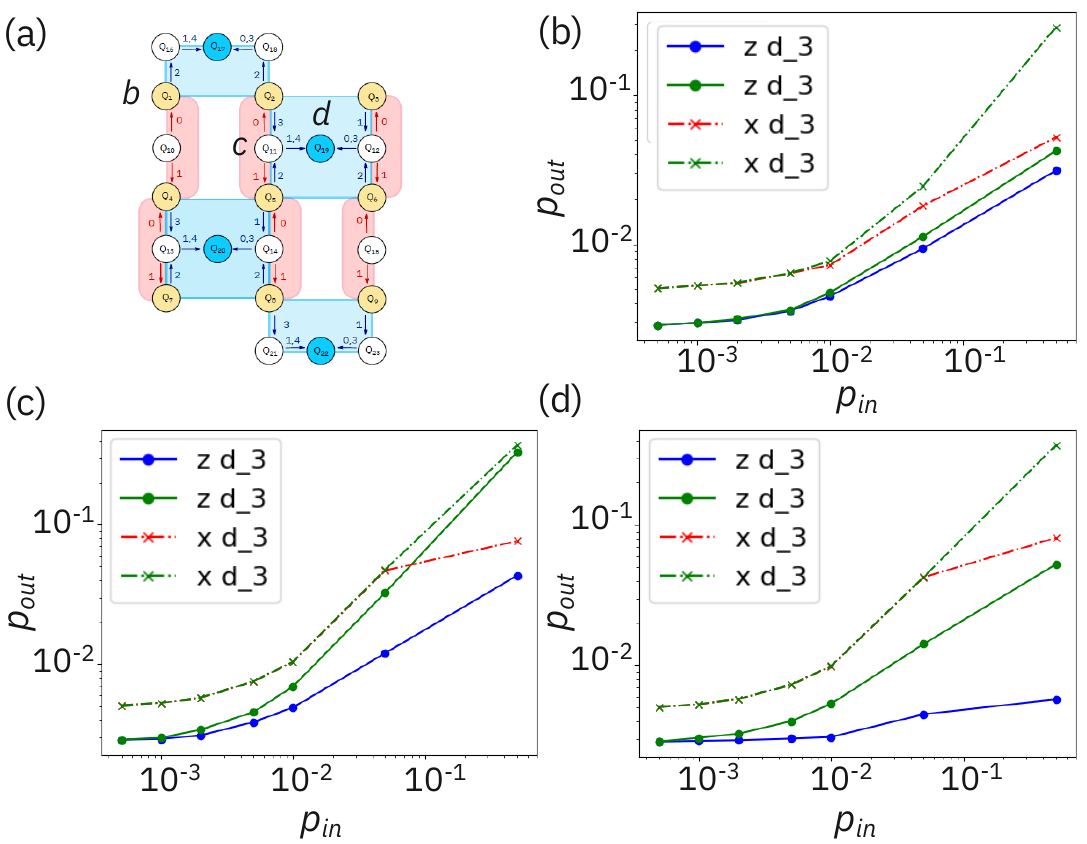}
    \caption{(a) The layout of the distance 3 heavy hex code with labelled sites, following the same notation as Sundaresan et al. \cite{Sundaresan_2023}. Z (blue) and X (red) gauge operators. (b) The logical output error rate dependence on $P_{in}$ of the \textit{cx} involved in operations with the qubit at the site labeled \textbf{b}, a data qubit. (c) Logical output error rate dependence on a flag qubit at the location marked \textbf{c} and the ancilla measure qubit for the Z check at the side marked \textbf{d}. The $\langle p_{in} \rangle = 5\times10^{-4}$ for all three figures. Error rates shown in green are for \textit{naive} decoding.}
    \label{fig:Fig4}
\end{figure*}

% We now check \textit{sensitivity} to other functional sites in the code (i.e., gauge and ancillae sites) and compare to the corner data qubit case for distances 3, 5 and 7, Fig. \ref{fig:Fig4}. 
In this appendix we illustrate the general trends and relative \textit{sensitivity} of 'bad' functional sites in a distance 3 HHC, Fig. \ref{fig:Fig4}. The S-shape behavior is generally repeated and the onset of increased output error is also qualitatively similar as described in the main text. That is, there is relatively small increase in $p_{out}$ until $p_{in}$ is large enough to begin to shift $\langle p_{in} \rangle$. We see, however, that there are differences in magnitude of \textit{sensitivity} and some sites are more or less sensitive to X or Z initialization and measurement, a consequence of their specific roles in the measurement of the stabilizers. We also show the difference in error rates for different decoding choices \textit{aware} (red or blue) and \textit{naive} (green).    

\section{Decoder notes}
\label{Appendix:Decoder}
A minimum weight perfect matching decoder was used for all results in this work. As described in Sundaresan et al. \cite{Sundaresan_2023} the perfect matching algorithm considers X and Z errors separate. The algorithm finds the minimum weight perfect matching in a graph to associate the syndrome with a particular error. A graph is formed of vertices representing error-sensitive events, V, and hyperedges representing the correlations between the events caused. The probabilistic circuit errors of each operation combine to form the correlation for each edge. There is a graph for each the X and Z errors. Edge weights are set as $w_e = \log((1-p_e)/p_e)$, where $p_e$ is the edge probability estimated from the leading order of the polynomial error rate associated with the parameterized gate operation error rates. The \textit{naive} decoder assumes that the error rates are the same for gates of the same kind of operation, while the \textit{aware} decoder uses the error rate that is actually at each site (e.g., selected from the random distribution). The decoder used in this work is the same as that used in Sundaresan et al. \cite{Sundaresan_2023} in which the details of the decoder are more extensively discussed.

\section{Complementary simulations for bad qubits with bias noise and distance dependences}

\begin{figure*}[]
    \centering
    \subfloat[\label{fig:dists:a}]{
    	\includegraphics[width=0.52\textwidth, clip,trim= 0mm 0mm 0mm 0mm ]{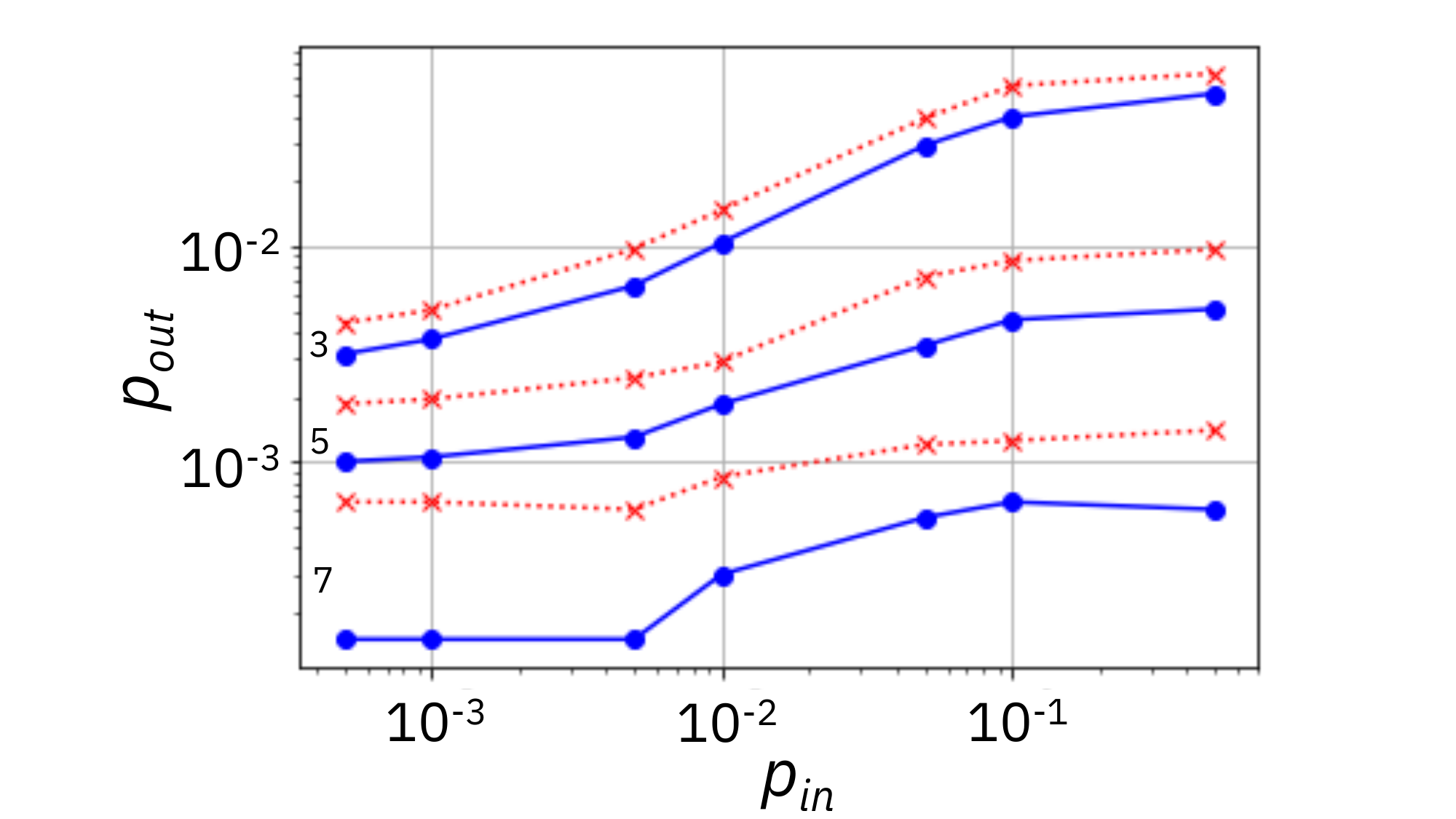}
    % 	\caption{histogram for $\alpha=0.3$}
    }
    \subfloat[\label{fig:ZZ:b}]{
        \includegraphics[width=0.5\textwidth, clip,trim= 0mm 0mm 0mm 0mm ]{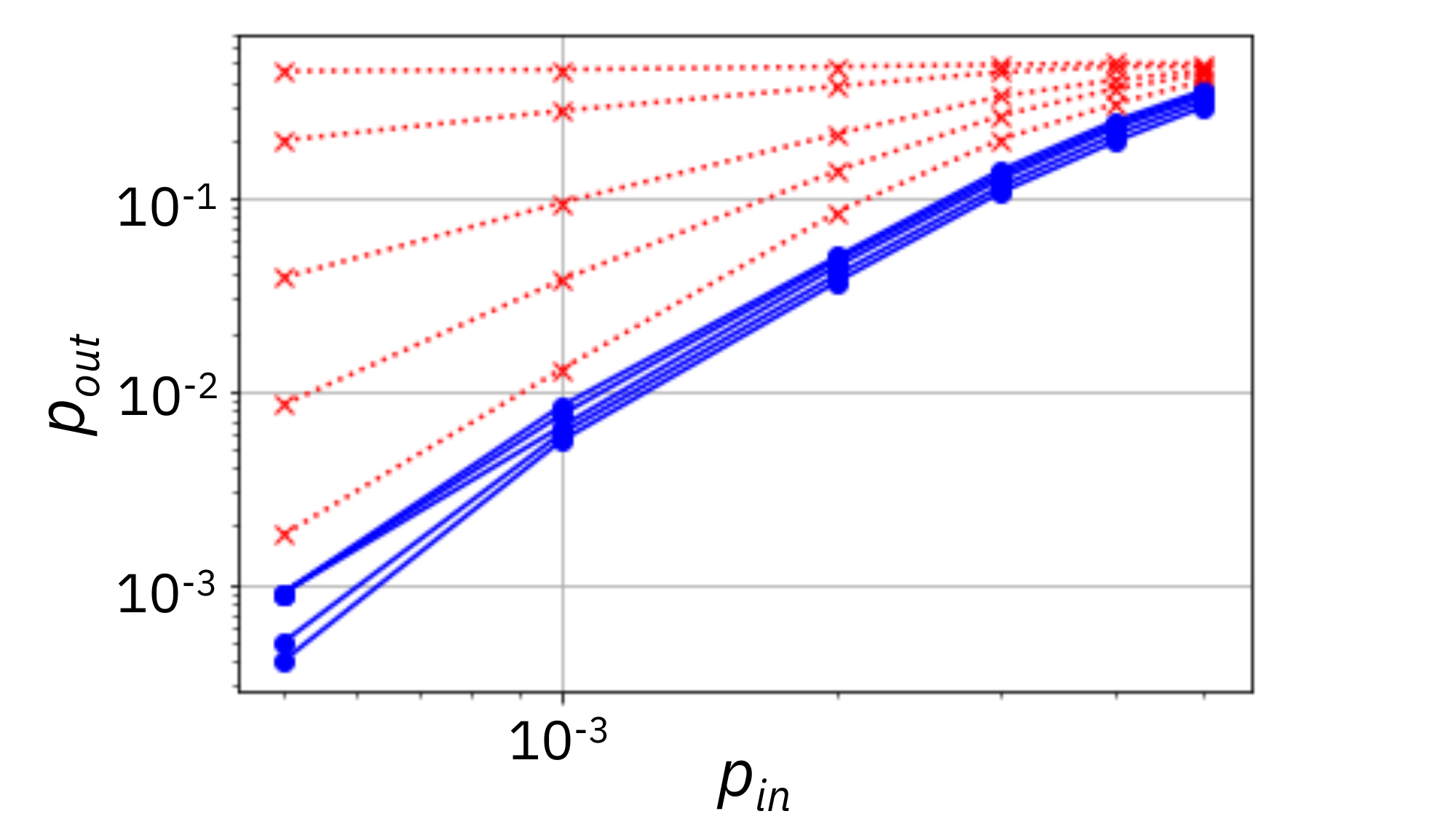}
        % \caption{mean dependence on $\alpha$}
    }

    \caption{(a) Output error rate dependence on a 'bad' qubit at site 0 (see Fig. \ref{fig:Fig1}(a)) for distances of 3, 5 and 7 using $\langle p_{in} \rangle = 5 \times 10^{-4}$. The number of shots were 5k, 5k and 10k, respectively. (b) Output error for distance 5 with IZ, ZI and ZZ set to $p_{in}=0.5$, 'bad', data qubits at positions 0, 1, 2, 3 and 4 (Fig. \ref{fig:Fig1} (a)). Red indicates X-measurement and blue indicates Z-measurement.}
    \label{fig:distsZZandXX}
\end{figure*}

In this appendix we illustrate the $p_{out}$ dependence on different aspects of high infidelity outlier error, complementing the example discussed in the main text. In figure \ref{fig:distsZZandXX} (a) we first show the distance dependence of $p_{out}$ when there is a single 'bad' qubit in the corner site 0. The $p_{in}$ of the 'bad' qubit is increased while the other errors are held constant as in the main text. One observation is that the error saturates in a qualitatively similar way for each distance.

In figure \ref{fig:distsZZandXX} (b) we examine the bias dependence by setting the error rate for ZI, ZZ and IZ to $p_{in}=0.5$ when linked to the 'bad' qubit site(s). The bad qubits are sequentially chosen along the Z logical operator. Depolarizing noise is still applied to all other locations. When the logical state is prepared and measured in the Z basis, it is insensitive to the additional 'bad' Z errors. Qualitatively, when preparing and measuring in the X basis, a similar weak response to XI, XX and IX error is correspondingly observed (not shown). The relative sensitivity is an example of the Z vs. X related $p_{out}$ response to biased 'bad' qubit noise.

\end{document}